\documentclass[useAMS,usenatbib]{mn2e}
\usepackage{graphicx}
\usepackage{epsfig}
\usepackage{amssymb}
\usepackage{lscape}
\usepackage{ulem}
\usepackage{txfonts}
\usepackage{lipsum}
\usepackage{float}
\usepackage{hyperref}
\usepackage[english]{babel}
\usepackage{blindtext}
\usepackage{aas_macros}
\usepackage[utf8]{inputenc}
\usepackage{longtable}

\title[Metallicity conversions] {A re-assessment of strong line metallicity conversions in the machine learning era.}

\author[ ]{Hossen Teimoorinia$^{1,2}$,
 Mansoureh Jalilkhany$^{3}$, Jillian M. Scudder$^{4}$ 
, 
Jaclyn Jensen $^{2}$, \newauthor Sara L. Ellison$^{2}$ 
\\ 
$^1$ NRC Herzberg Astronomy and Astrophysics, 5071 West Saanich Road, Victoria, BC, V9E 2E7, Canada \\
$^2$ Department of Physics and Astronomy, University of Victoria, Victoria, BC, V8P 5C2, Canada\\ 
$^3$ Department of Computer Science, University of Victoria, Victoria, BC, V8W 2Y2, Canada\\
$^4$ Department of Physics and Astronomy, Oberlin College, Oberlin, Ohio, OH 44074, USA}

\def\LaTeX{L\kern-.36em\raise.3ex\hbox{a}\kern-.15em
    T\kern-.1667em\lower.7ex\hbox{E}\kern-.125emX}

\begin{document}

\label{firstpage}

\maketitle

\begin{abstract} 

 Strong line metallicity calibrations are widely used to determine the gas phase metallicities of individual HII regions and entire galaxies.  Over a decade ago, based on the Sloan Digital Sky Survey Data Release 4 (SDSS DR4), Kewley \& Ellison published the coefficients of third-order polynomials that can be used to convert between different strong line metallicity calibrations for global galaxy spectra.  Here, we update the work of Kewley \& Ellison in three ways.  First, by using a newer data release (DR7), we approximately double the number of galaxies used in polynomial fits, providing statistically improved polynomial coefficients.  Second, we include in the calibration suite five additional metallicity diagnostics that have been proposed in the last decade and were not included by Kewley \& Ellison.  Finally, we develop a new machine learning approach for converting between metallicity calibrations.  The random forest algorithm is non-parametric and therefore more flexible than polynomial conversions, due to its ability to capture non-linear behaviour in the data.  The random forest method yields the same accuracy as the (updated) polynomial conversions, but has the significant advantage that a single model can be applied over a wide range of metallicities, without the need to distinguish upper and lower branches in $R_{23}$ calibrations.  The trained random forest is made publicly available for use in the community.
  
\end{abstract}


\section{Introduction}
\label{introduction}

The gas phase metallicity of a galaxy is a fundamental quantity that captures the cumulative history of stellar chemical enrichment, outflows and delivery of pristine material from the intergalactic medium.  Commonly denoted as 12 + log(O/H), this metallicity is encoded into the emission line spectrum of an HII region or galaxy, and can be derived `directly' by solving for the electron temperature and density.  Electron densities are readily determined from nebular doublets, such as [OII] $\lambda \lambda$ 3726, 3729.  In theory, electron temperatures ($T_e$) are similarly readily obtained, by measuring the relative strengths of emission lines of a given species that originate from different upper energy levels, e.g. [OIII] $\lambda$ 5007 and the auroral line [OIII] $\lambda$ 4363.  

However, in practice, the weakness of the auroral lines that are used to measure $T_e$ limits this method to metallicities below approximately 12 + log (O/H) $\sim$ 8.3 for most individual spectra.   The auroral line becomes weaker as metallicity increases due to the increased cooling in the interstellar medium.  A common alternative practice is therefore to use `strong line methods' which use readily detectable nebular emission line ratios that have been calibrated against `known' metallicities \citep[see][for a review]{Kewley-19}.  A large variety of strong line metallicity calibrations exists in the literature, broadly falling into two categories; the so-called `empirical' methods, that are calibrated against direct ($T_e$) metallicity measurements \citep[e.g.,][]{Pettini-04,Marino-13,Curti-17} and those that are calibrated against theoretical models \citep[e.g.,][]{McGaugh-91,Zaritsky-94,Kewley-02} 

Although the strong line metallicity calibrations show good internal consistency, there are strong systematic offsets between them.  Therefore, attempting to combine published measurements of gas phase metallicity that have used different metallicity calibrations will result in error, with offsets as large as 0.7 dex \citep[e.g.,][]{Kewley-08}.  In order to facilitate the combination of metallicities derived using different methods, \cite{Kewley-08} published an extensive assessment of the major strong line metallicity diagnostics of the time.  Using metallicity data for $\sim$ 28,000 galaxies in the Data Release 4 (DR4) of the Sloan Digital Sky Survey (SDSS), \cite{Kewley-08} published tables of coefficients for third-order polynomials that could be used to convert between any pair of diagnostics used in their study.

Although the metallicity conversions of \cite{Kewley-08} continue to be widely used, in the decade since their publication several advances have been made in the community.  First, the samples of star-forming galaxies that can be used to calibrate metallicities has grown considerably.  Secondly, several new metallicity calibrations have been introduced in the literature for which no conversions (to other diagnostics) exist.  Finally, the rapid growth of machine learning technologies means that new methods exist for calibrating between different diagnostics.  Such methods offer numerous advantages over the polynomial fitting approach of \cite{Kewley-08}.  For example, they can capture non-linearities in the data, make no assumptions on the form of the conversion ($viz$ the third-order polynomial used in Kewley \& Ellison 2008) and simplify the process by avoiding different functional forms in different regimes (e.g. upper and lower metallicity branches).

Machine learning has already gained a solid foothold in astronomical data processing.  For example, neural networks and deep learning methods have been widely used for both image classification \citep{Bottrell-19,Huertas-Company-19,Ciprijanovic-20,Ferreira-20,Teimoorinia-20b,Teimoorinia-20a}, ranking tasks \citep{Teimoorinia-16,Bluck-19,Dey-19,Ellison-20}, as well as for regression and pattern recognition applications \citep{Ellison-16,Teimoorinia-17}.  In the realm of interstellar medium studies, neural networks and other techniques, such as random forests, have also been used to predict emission line fluxes \citep{Teimoorinia-14} and metallicities from either broad band photometry \citep{Acquaviva-15}   or spectra \citep{Ucci-18,Ho-19}. 

Given the advances made since the original work of \cite{Kewley-08}, the goal of the work presented here is three-fold.  First, using a larger sample of star-forming galaxies drawn from the SDSS Data Release 7\footnote{Despite its long presence in the literature, the DR7 remains one of the most widely used galaxy datasets for spectroscopic work in the nearby universe, thanks to the public MPA/JHU database of physical properties, including emission line fluxes.} (DR7; Sec.\ref{sec:data}), in Sec. \ref{sec:poly_method} we re-assess the polynomial-based metallicity conversions presented by\cite{Kewley-08}.  We show that there are small systematic errors (up to $\sim$ 0.1 dex) when the old \cite{Kewley-08} conversions are applied to the newer DR7 sample.  New coefficients are tabulated, providing a statistically improved set of functions for researchers desiring to use polynomial based conversions between metallicity diagnostics (Sec. \ref{sec:poly_ke08}).  Second, we extend the original suite of metallicity diagnostics with a further five calibrations that have been presented in the literature since the original work of \cite{Kewley-08}.  Finally, in Sec. \ref{sec:RF}, we develop a random forest alternative to the polynomial conversion approach.  We emphasize that our goal is not to develop new metallicity calibrations, but rather to facilitate the conversion between existing methods using modern samples and methods.  Our results are summarized in Sec. \ref{sec:summary}.

\section{Data}
\label{sec:data}

\subsection{A brief review of Kewley \& Ellison (2008)}
\label{sec:ke08_review}

We refer the reader to more extensive treatises on metallicity measurements in galaxies for a full pedagogical description and theoretical framework of this field \citep[e.g.,][]{Pagel-89,Osterbrock-89,Stasinska-06,Lopez-Sanchez-12,Perez-Montero-17,Kewley-19}.  Here, we review only the essential concepts, data selection and metallicity calibrations used in \cite{Kewley-08} as context for our re-assessment of that work.

Strong line metallicity diagnostics combine the fluxes from two or more nebular emission lines, which can be input into (usually, simple polynomial) functions that have been either empirically or theoretically calibrated in order to solve for the gas phase metallicity, O/H.  The function that converts line fluxes to O/H are not always monotonic.  One of the most well known examples of a strong line diagnostic that is double valued in O/H for a given line ratio, is the family of $R_{23}$ diagnostics, where

\begin{equation}
  R_{23} = \frac{[OII] \lambda 3727 + [OIII] \lambda 4959 + [OIII] \lambda 5007}{H\beta}
\end{equation}

Additional constraints are required to break the $R_{23}$ degeneracy in order to determine whether the galaxy (or HII region) lies on the upper or lower branch of the calibration.  \cite{Kewley-08} use the ratio of [NII]/[OII] to break the $R_{23}$ degeneracy, since this ratio is not sensitive to ionization parameter and largely monotonic with metallicity in the range of SDSS galaxy metallicities.  Whilst this approach can work well for metal-rich and metal-poor galaxies, at intermediate metallicities the calibration can be ambiguous and O/H is more uncertain.

\cite{Kewley-08} select their galaxy sample from the SDSS DR4.  A S/N of at least 8 is required in the [OII], [OIII], [NII], [SII] and Balmer lines used in the metallicity calibrations.  $g$-band fibre covering fractions are required to be at least 20 per cent, enforcing an effective lower redshift cut $z \sim 0.04$. An upper redshift cut of $z<0.1$ is also imposed.  Galaxies dominated by active galactic nucleus (AGN) emission are removed using the criteria presented in \cite{Kewley-06}.  The resulting sample contains $\sim$ 28,000 star-forming galaxies.  Emission line fluxes were corrected by measuring the Balmer decrement and applying the Milky Way extinction curve of \cite{Cardelli-89}.

\medskip

\cite{Kewley-08} considered 10 different metallicity calibrations: the direct (or $T_e$) method and nine different strong line methods.  $T_e$ based metallicities are not available for the majority of individual SDSS galaxies, hence we consider it no further here (but see \citep{Andrews-13,Curti-17} for works that have derived $T_e$ based metallicities for SDSS spectral stacks).  Of the nine strong line methods, we exclude two from our current work.  First, we exclude the calibration of \cite{Denicol-02}; this calibration is based on the ratio of [NII]/H$\alpha$ and has been superseded by several other calibrations used in this work. Second, we also exclude \cite{Pilyugin-05}; as shown by Kewley \& Ellison (2008) this diagnostic does not cross-calibrate well with other methods.

Below we briefly review the remaining seven strong line methods used by Kewley \& Ellison (2008) that we will re-calibrate in the work presented here:

\begin{itemize}

\item \cite{McGaugh-91}.  Hereafter M91, this calibration uses $R_{23}$ and is calibrated using theoretical models.  \cite{Kewley-08} calibrate the upper and lower branches of this diagnostic with separate polynomial fits, whose degeneracy is broken with the [NII]/[OII] ratio.
  
\item \cite{Zaritsky-94}.   Hereafter Z94, this is an average of three previously published theoretically calibrated $R_{23}$ diagnostics and is applicable only to the upper branch.
 
\item \cite{Kewley-02}.   Hereafter KD02, this diagnostic is theoretically calibrated.  KD02 is applied in two regimes - at high metallicities the calibration is based on the ratio [NII]/[OII], and at low metallicities it uses an average of several $R_{23}$ methods.

\item \cite{Kobulnicky-04}.   Hereafter KK04, this calibration uses the same theoretical grids as KD02.  Application of the KK04 calibration is a multi-step process that entails an assessment of whether the galaxy is on the upper or lower $R_{23}$ branch, following an initial estimate of the ionization parameter from the ratio of [OIII]/[OII].  The final metallicity and ionization parameter is determined through an iterative process based on $R_{23}$.
   
\item \cite{Pettini-04} N2 and O3N2.  Hereafter PP04 N2 and PP04 O3N2, these diagnostics are both empirically\footnote{A small number (6/137) of the HII regions used in the Pettini \& Pagel (2004) diagnostic are calibrated using theoretical models.  Given the vast majority of $T_e$ based metallicities used by PP04, we refer to this as an empirical method.} calibrated against direct ($T_e$) metallicities, and  use either the ratio of [NII]/H$\alpha$ (PP04 N2) or additionally including [OIII]/H$\alpha$ (PP04 O3N2).  Specifically, the following indices are defined:

\begin{equation}
  N2 = \rm{log[([NII]\lambda6583/H\alpha)]}
\end{equation}

and

\begin{equation}
  O3N2 = \rm{log[([OIII]\lambda5007/H\beta)/([NII]\lambda6583/H\alpha)]}
\end{equation}

  The advantage of both of N2 and O3N2 calibrations is that they are single valued and do not require dust correction of the emission line fluxes, because the lines used in the ratios are close in wavelength.  However, PP04 N2 (as with any N2 based calibration) has the limitation that it tends to saturate at higher metallicities.

 \item \cite{Tremonti-04}.   Hereafter T04, this calibration uses a broader suite of emission lines, including [NII] and [SII] in addition to the commonly used oxygen and Balmer lines.  Unlike the other calibrations used by \cite{Kewley-08}, T04 derive metallicities from a statistical assessment of a large suite of spectral synthesis models.   The T04 metallicities are therefore not computed by us (nor were they by Kewley \& Ellison 2008) from the raw emission line fluxes, but rather are taken directly from the MPA/JHU catalog.  We note that the large number of galaxies with available T04 metallicities, despite the large number of lines required, is because of the specific S/N requirements adopted in that work (which are not applied to all of the listed lines). 
\end{itemize}

\subsection{New calibrations considered in this paper}\label{sec:new_calibs}

We include five new metallicity calibrations in the current study that were not available to \cite{Kewley-08}, but have become widely used in the literature since their publication.  We review here their main properties, for comparison with the diagnostics used in \cite{Kewley-08}, but refer the reader to the original papers for full details.

\medskip

\begin{itemize}

\item \cite{Marino-13} N2 and O3N2.  Hereafter M13 N2 and M13 O3N2, this study compiled over 600 HII regions with $T_e$ based metallicities.  Two strong line diagnostics were calibrated based upon this large sample, [NII]/H$\alpha$ (i.e. the N2 index) and ([OIII]/H$\beta$)/([NII]/H$\alpha$) (i.e. the O3N2 index).  As with other N2 diagnostics, saturation is a problem at higher metallicities, whereas O3N2 can be used up to 12 + log(O/H) $\sim$ 8.9.

\item \cite{Dopita-16}.  Hereafter D16, this diagnostic uses a unique set of emission lines amongst those studied here.  By combining [NII], H$\alpha$ and [SII], which are all located close together in wavelength space, this metallicity diagnostic offers several advantages.  First, it is independent of extinction.  Second, all of the emission lines can be covered in one spectral setting, even if the instrumental coverage is relatively narrow.  Finally, this diagnostic is characterised by a linear function up to super-solar metallicities (12+log(O/H) $\approx$ 9.05) and is largely independent of ionization parameter.

\item\cite{Curti-17} N2 and O3N2.  Hereafter C17 N2 and C17 O3N2, this study used over 110,000 star-forming galaxies drawn from the SDSS DR7 (the same dataset from which we will draw our sample) to derive an empirical metallicity calibration over an unprecedentedly broad range in O/H by stacking galaxies in narrow bins of [OII] and [OIII] relative to H$\beta$.  The very high S/N ratios achieved in the stacking process enable a detection of [OIII] $\lambda$ 4363 , which is required to determine the electron temperature and thereby solve for metallicity.  As with \cite{Marino-13}, \cite{Curti-17} calibrated their $T_e$ based metallicities for two strong line ratios, N2 and O3N2. 
  
\end{itemize}

Table \ref{tab:metal-ref} summarizes the 12 diagnostics used in this paper; seven from the original \cite{Kewley-08} work and five new ones.    Table \ref{tab:metal-ref} also summarizes the metallicity regime over which the calibrations are valid and the emission lines used.

\subsection{SDSS DR7 Galaxy Selection}

\begin{table*}
\caption{The list of metallicity calibrations used in this paper and number of galaxies in the DR7 dataset.}
\begin{tabular}{lcccc}
\hline
Metallicity & Reference &Range of validity & Lines required & \textit{N} galaxies  \\
\hline

$\rm{M91}$ & $\rm{McGaugh ~(1991)}$ & $\rm{7.24 <12+ log(O/H) <9.4}$ & $\rm{[NII]\lambda6583, [OII]\lambda3727,  [OIII]\lambda5007, H\beta}$ & 32275  \\ 

$\rm{Z94}$ & $\rm{Zaritsky~ et~ al. ~(1994)}$ & 12+log(O/H)$>$ 8.35 &$\rm{[NII]\lambda6583, [OII]\lambda3727, [OIII]\lambda5007, H\beta}$  & 31701 \\ 

$\rm{KD02}$ & $\rm{Kewley ~\& ~Dopita~ (2002)}$ & $\rm{8.2 <12+ log(O/H) < 9.4}$ & $\rm{[NII]\lambda6583, [OII]\lambda3727, [OIII]\lambda5007, H\beta}$ & 37899  \\ 

$\rm{KK04}$ & $\rm{Kobulnicky ~\& ~Kewley ~(2004)}$ & $\rm{7.6 <12+ log(O/H) < 9.2}$ & $\rm{[NII]\lambda6583, [OII]\lambda3727, [OIII]\lambda5007, H\beta}$ &  32168 \\

$\rm{PP04~N2}$ & $\rm{Pettini ~\& ~Pagel~ (2004)}$ & $\rm{-2.5 < N2 < -0.3
}$ & $\rm{[NII]\lambda6583, H\alpha}$ & 44726 \\ 

$\rm{PP04~O3N2}$ & $\rm{Pettini ~\&~ Pagel ~(2004)}$ & $\rm{O3N2 < 2.0}$ & $\rm{[OIII]\lambda5007, H\beta, H\alpha, [NII]\lambda6583}$ & 44590 \\ 

$\rm{T04}$ & $\rm{Tremonti~et~al. ~(2004)}$ &  N/A & $\rm{[NII]\lambda6583, [OII]\lambda3727,  [OIII]\lambda5007, [OI]}$  &  50774 \\ 

 &  & & $\rm{HeI, H\beta, H\alpha, [SII] \lambda\lambda6717, 6731}$  &   \\

$\rm{M13~N2}$ & $\rm{Marino~ et ~al. ~(2013)}$ & $\rm{-1.6 < N2 < -0.2}$ & $\rm{[NII]\lambda6583, H\alpha}$ & 45036\\

$\rm{M13~O3N2}$ & $\rm{Marino ~et ~al. ~(2013)}$ & $\rm{-1.1 <O3N2< 1.7}$ & $\rm{[OIII]\lambda5007, H\beta, H\alpha, [NII]\lambda6583}$ & 44085\\ 

$\rm{D16}$ & $\rm{Dopita~ et ~al. ~(2016)}$ & $\rm{7.76 <12+ log(O/H) < 9.05}$ & $\rm{[NII]\lambda6583, H\alpha, [SII] \lambda\lambda6717, 6731}$ & 42892  \\ 

$\rm{C17~N2}$ & $\rm{Curti~ et~ al.~ (2016)}$ & $\rm{7.6 <12+ log(O/H) < 8.85}$ & $\rm{[NII]\lambda6583, H\alpha}$ & 43743   \\ 
$\rm{C17~O3N2}$ & $\rm{Curti~ et ~al.~ (2016)}$ & $\rm{7.6 <12+ log(O/H) < 8.85}$ &$\rm{[OIII]\lambda5007, H\beta, H\alpha, [NII]\lambda6583}$  &  44946 \\

\hline
\end{tabular}
\label{tab:metal-ref}
\end{table*}

\begin{figure*}
\centering
\includegraphics[width=5.2cm]{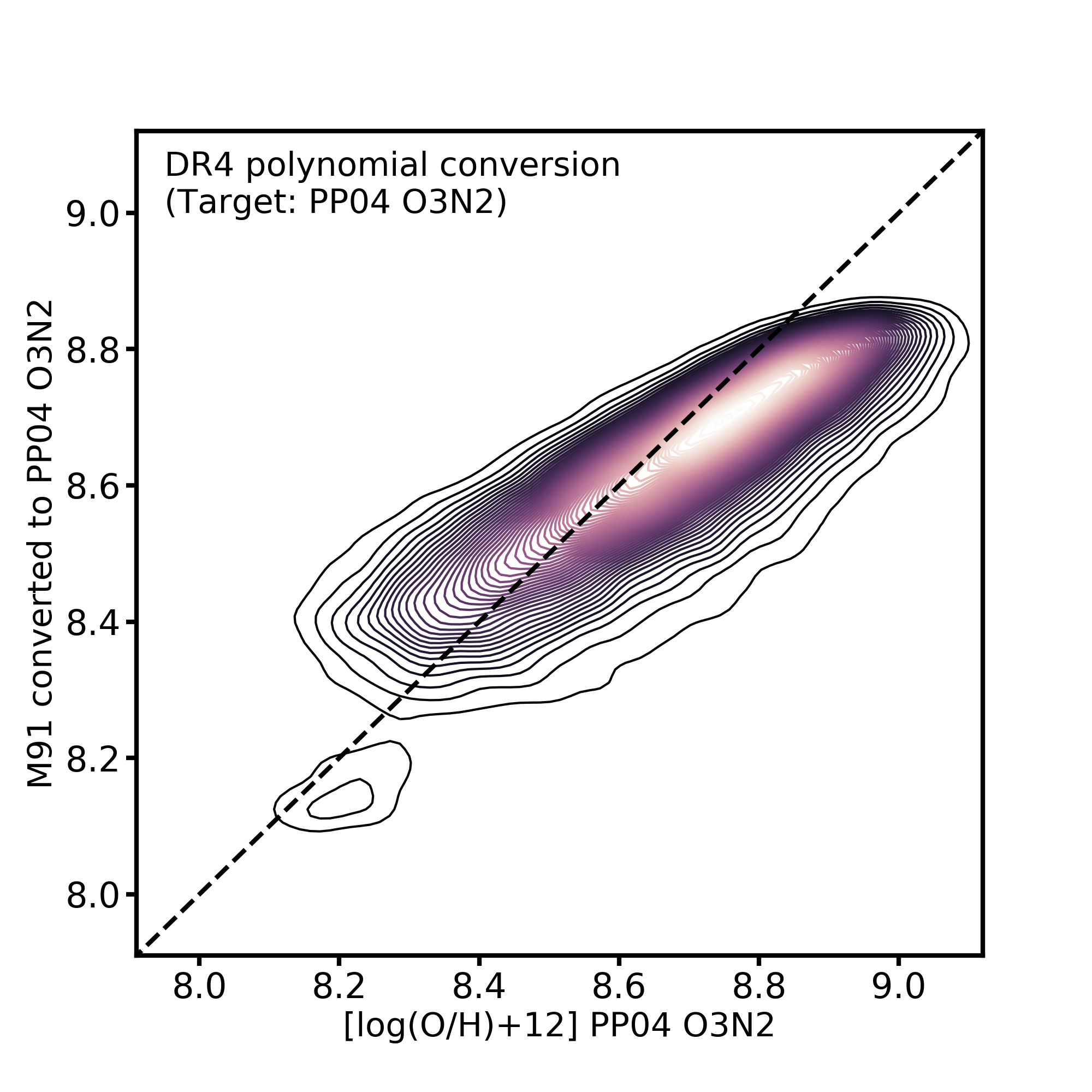}
\includegraphics[width=5.2cm]{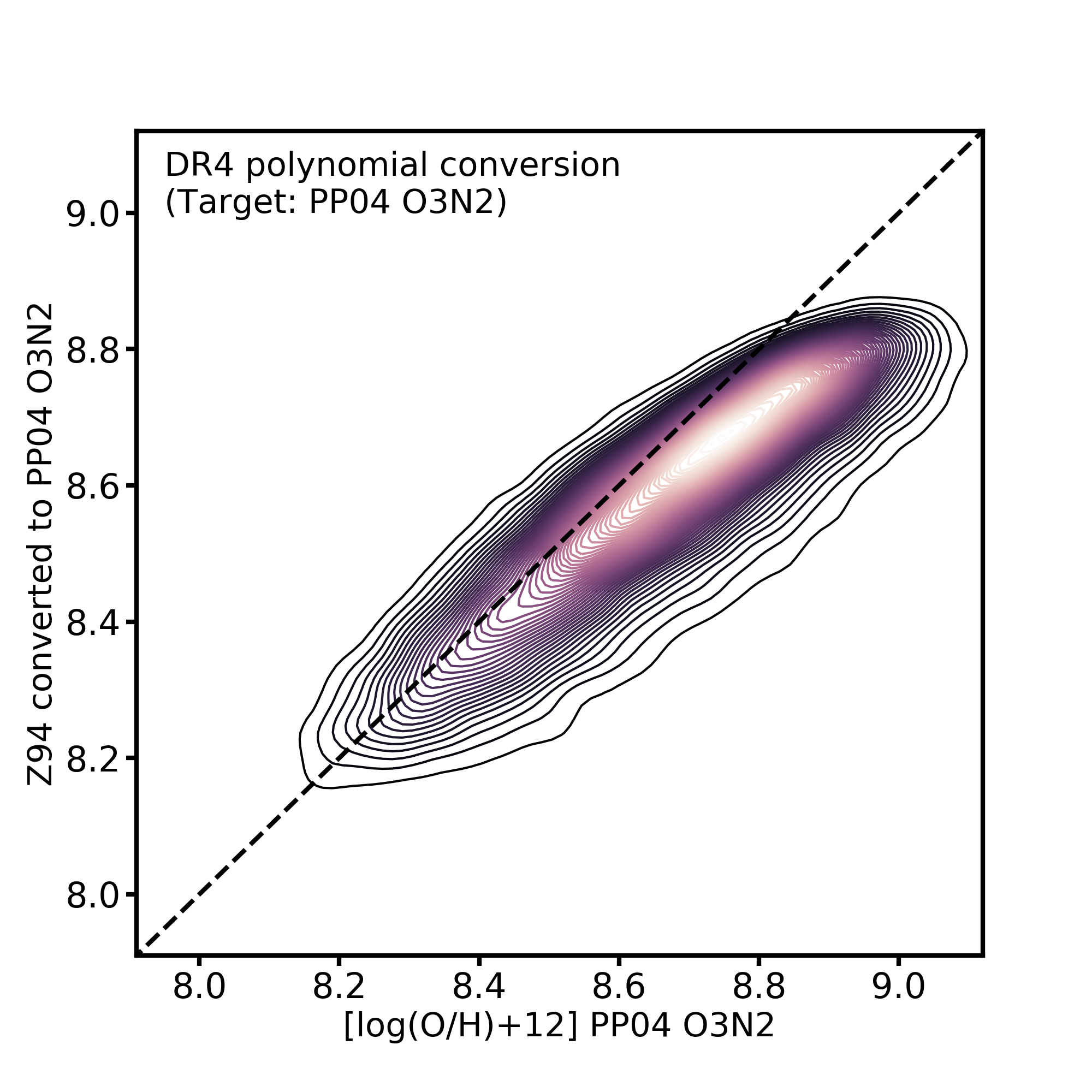}
\includegraphics[width=5.2cm]{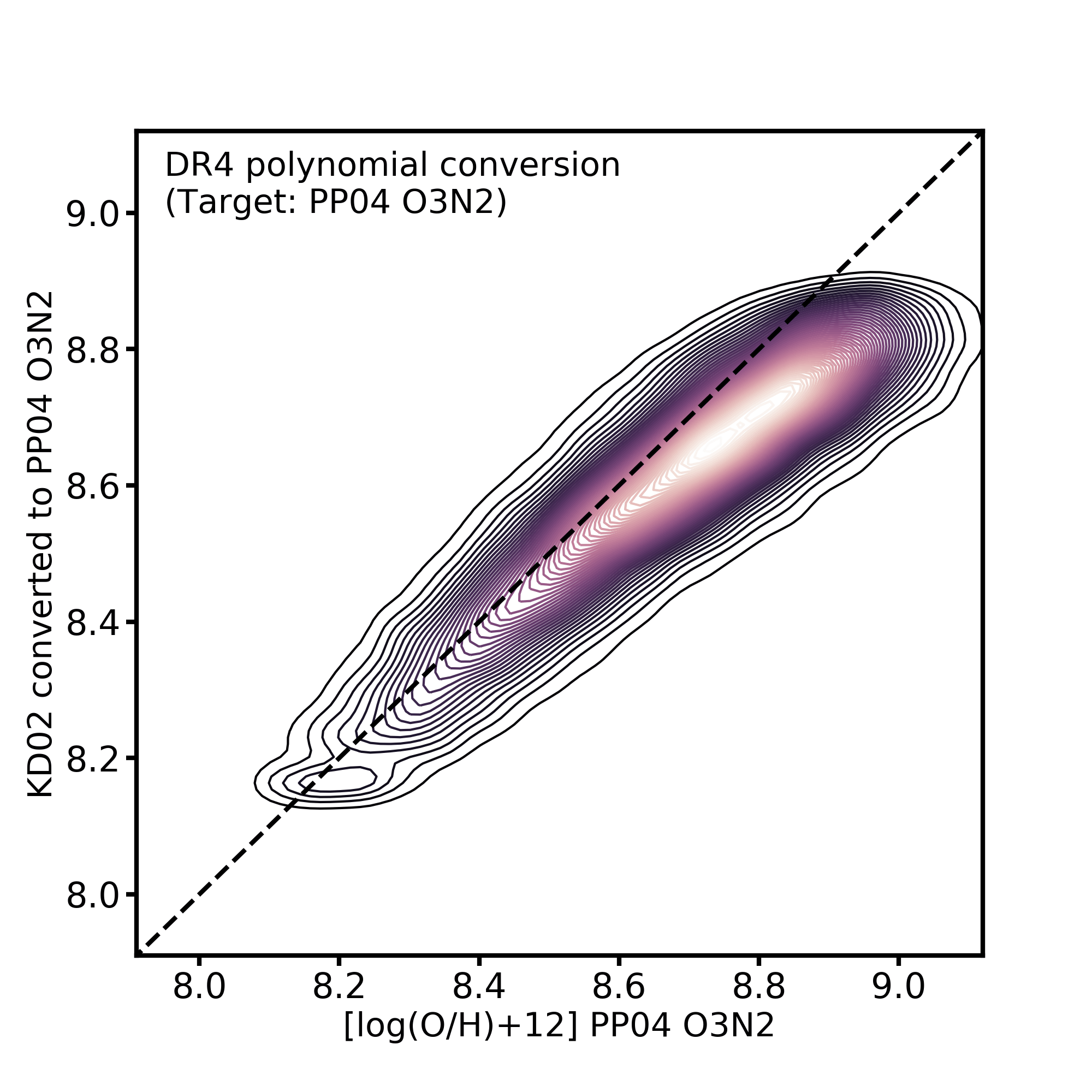}
\includegraphics[width=5.2cm]{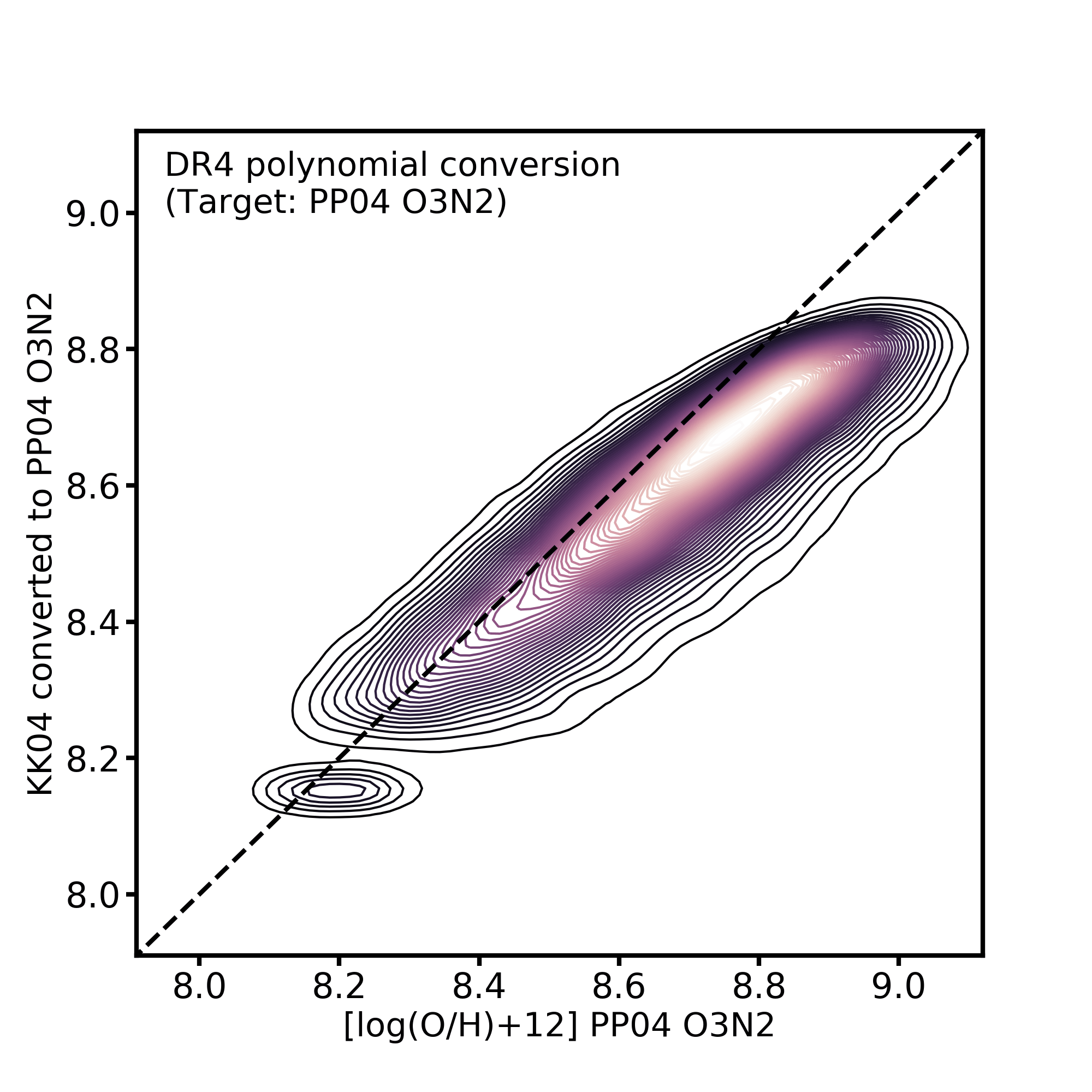}
\includegraphics[width=5.2cm]{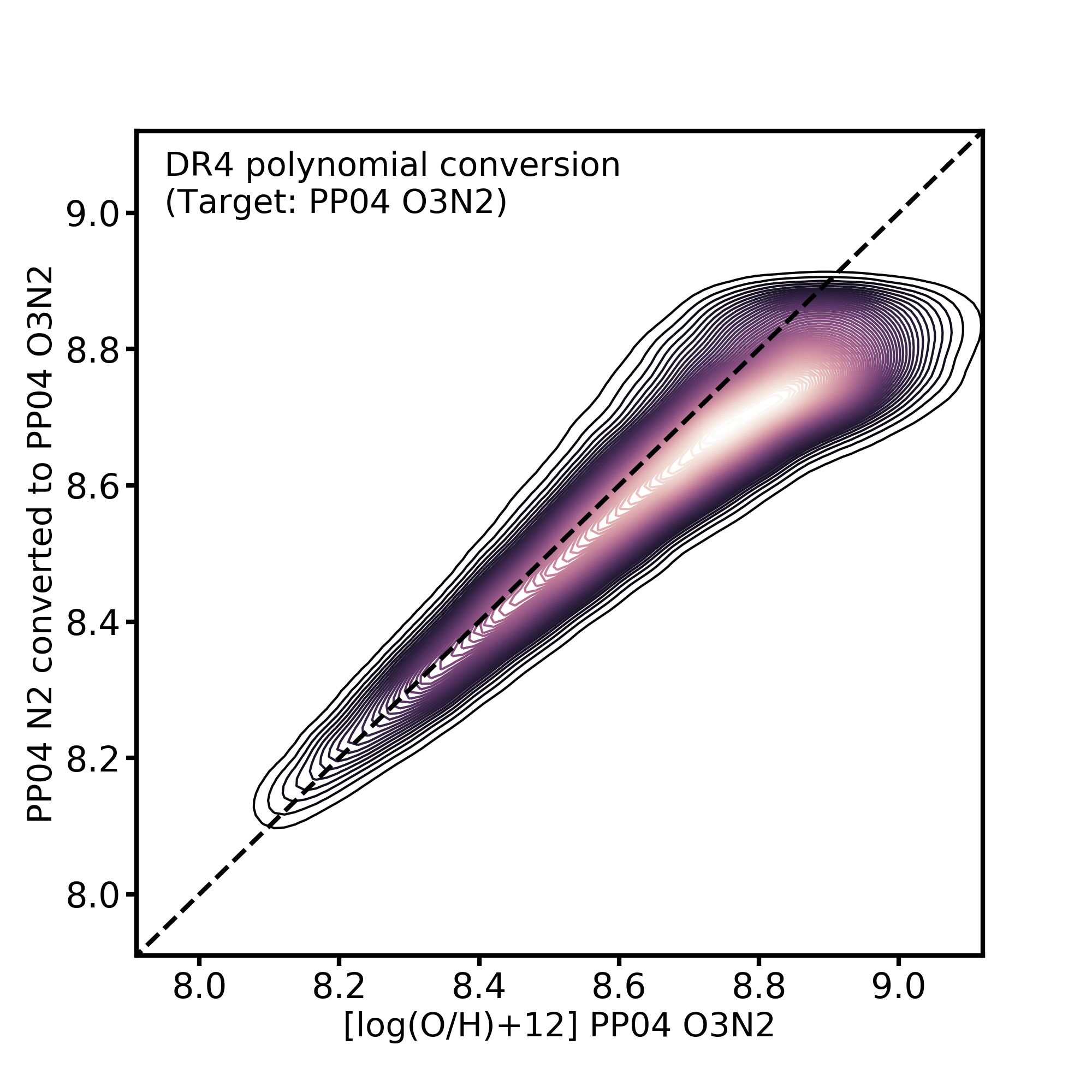}
\includegraphics[width=5.2cm]{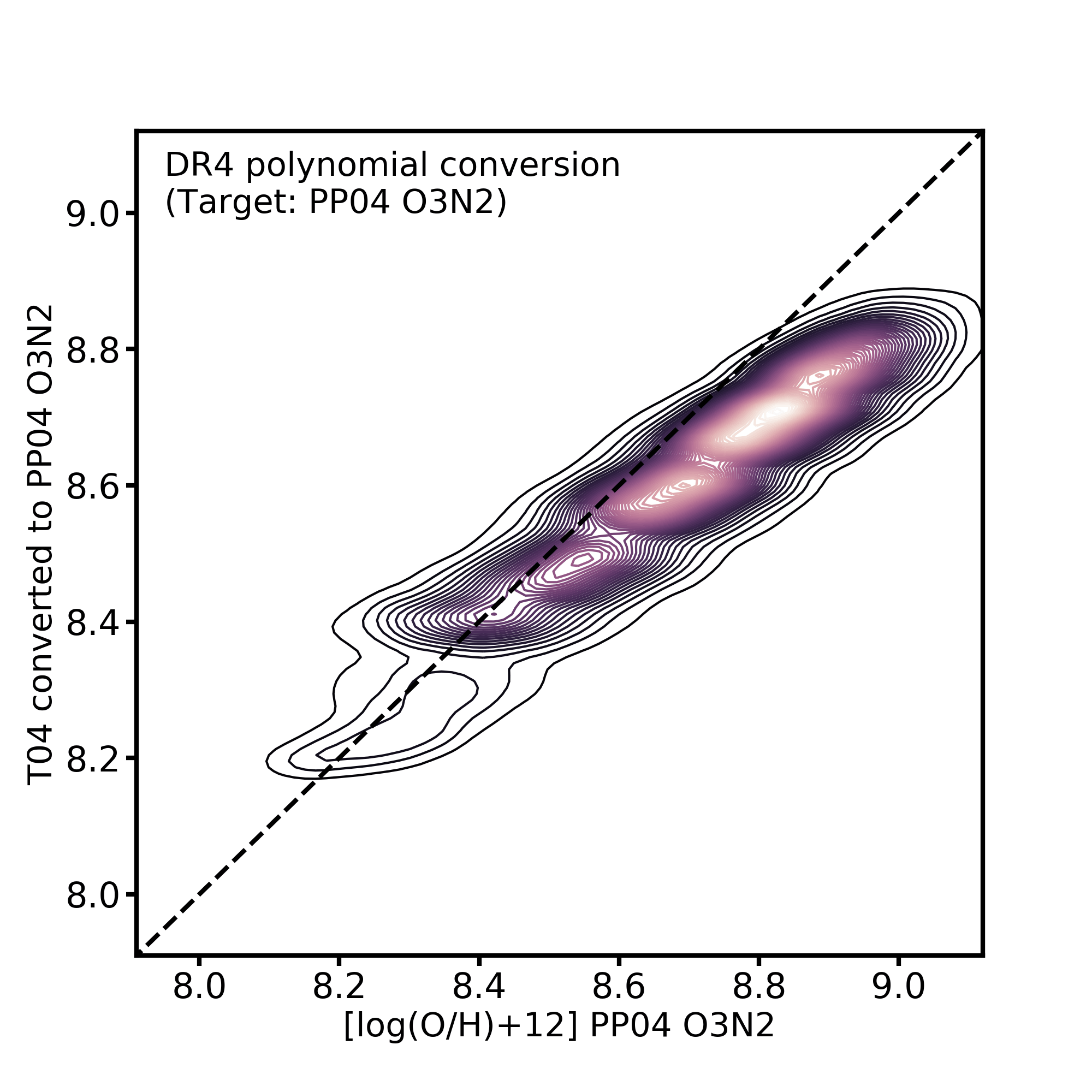}
\caption{Assessment of the DR4-derived metallicity conversions on the expanded DR7 dataset.  In each panel, the DR7-derived PP04 O3N2 metallicity is shown on the x-axis.  For the six other strong line methods calibrations with conversions by Kewley \& Ellison (2008) that are used in this paper, the y-axis shows each metallicity diagnostic converted to PP04 O3N2 using the original DR4 polynomial coefficients.  The deviations of the data from the diagonal 1:1 dashed line demonstrate that the DR4 conversions are not optimized for the DR7 dataset.}
\label{fig-old}
\end{figure*}

Emission line fluxes used in this work are taken from the publicly available SDSS DR7 MPA/JHU catalog\footnote{https://wwwmpa.mpa-garching.mpg.de/SDSS/}, e.g. \cite{Brinchmann-04}.  The emission lines in this catalog have been corrected for underlying stellar absorption and for Galactic extinction.  We further correct for internal extinction by assuming an intrinsic H$\alpha$/H$\beta$=2.85 and a Small Magellanic Cloud extinction curve (Pei 1992)\footnote{We test the influence of choosing a Milky Way type extinction curve and find that it has negligible impact on the calculated metallicities.}. 

We begin by selecting all galaxies from the DR7 that are classified as star-forming according to the \cite{Kauffmann-03} definition.  For this selection, we require that the S/N in the four emission lines required for the star-forming classification have a S/N $>$ 3.  Approximately 159,000 galaxies are thus selected.

In order to identify galaxies for which robust, global metallicities may be estimated, further cuts are required on emission line S/N and fibre covering fraction.  In order to facilitate an equitable comparison with the conversions of \cite{Kewley-08}, we adopt the same criteria as in that work, i.e., $g$-band fibre covering fraction of at least 20 per cent, an upper redshift cut of 0.1 and a minimum S/N in the emission lines required for a given calibration of 8.  The application of the first two of these criteria result in a sample that is reduced from $\sim$ 159,000 to $\sim$ 61,000.  The S/N requirement further reduces the sample to a size that depends on the specific set of emission lines required for a given calibration.  The final column of Table \ref{tab:metal-ref} summarizes the number of galaxies selected for each calibration.  The number of galaxies used in each pairwise cross-calibration (which requires a S/N $>$ 8 in the lines used for both calibrations in a given conversion) is further given in Tables \ref{tab:M91} - \ref{tab:C17-O3N2}.

\section{The polynomial conversion method}
\label{sec:poly_method}

\begin{figure*}
\centering
\includegraphics[width=5.2cm]{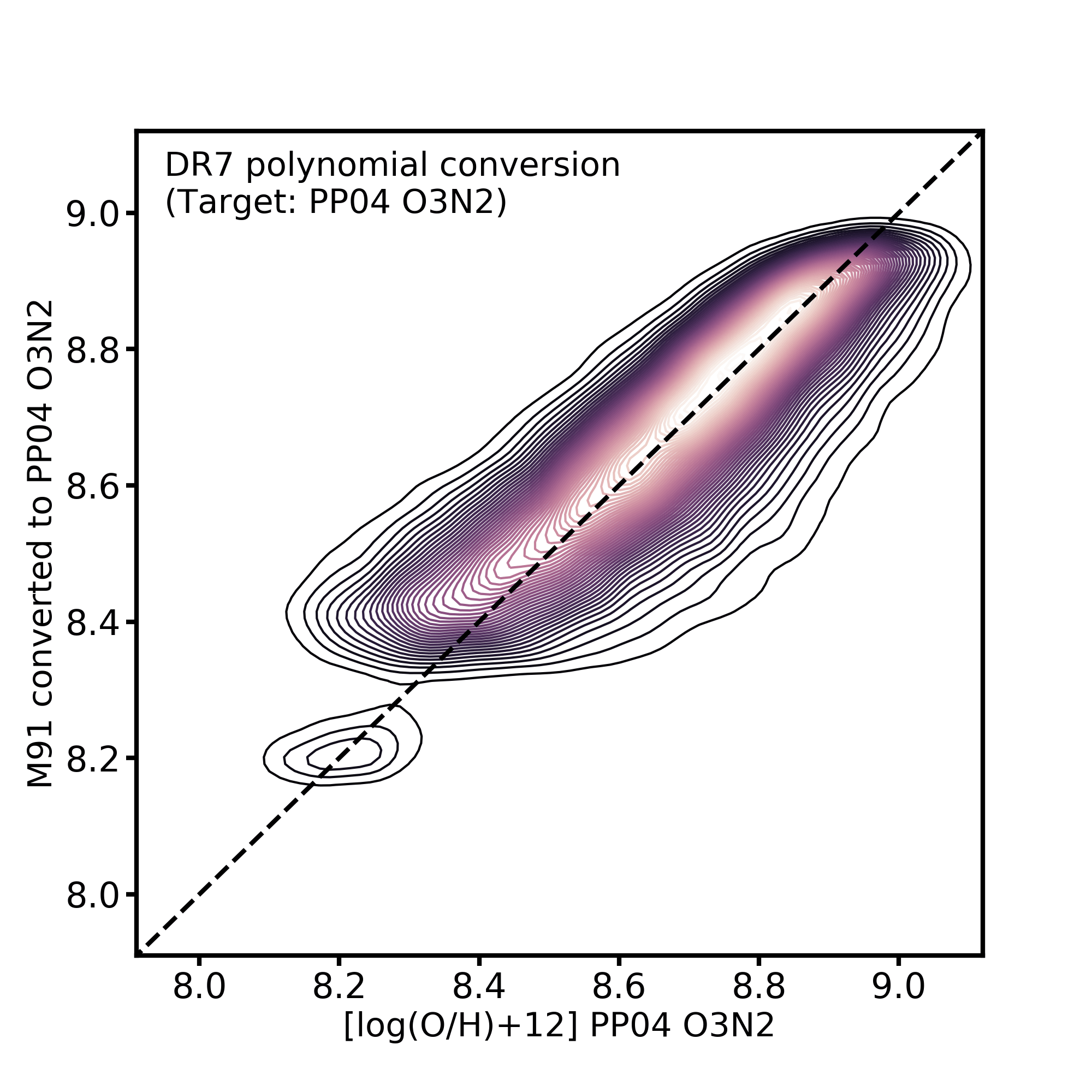}
\includegraphics[width=5.2cm]{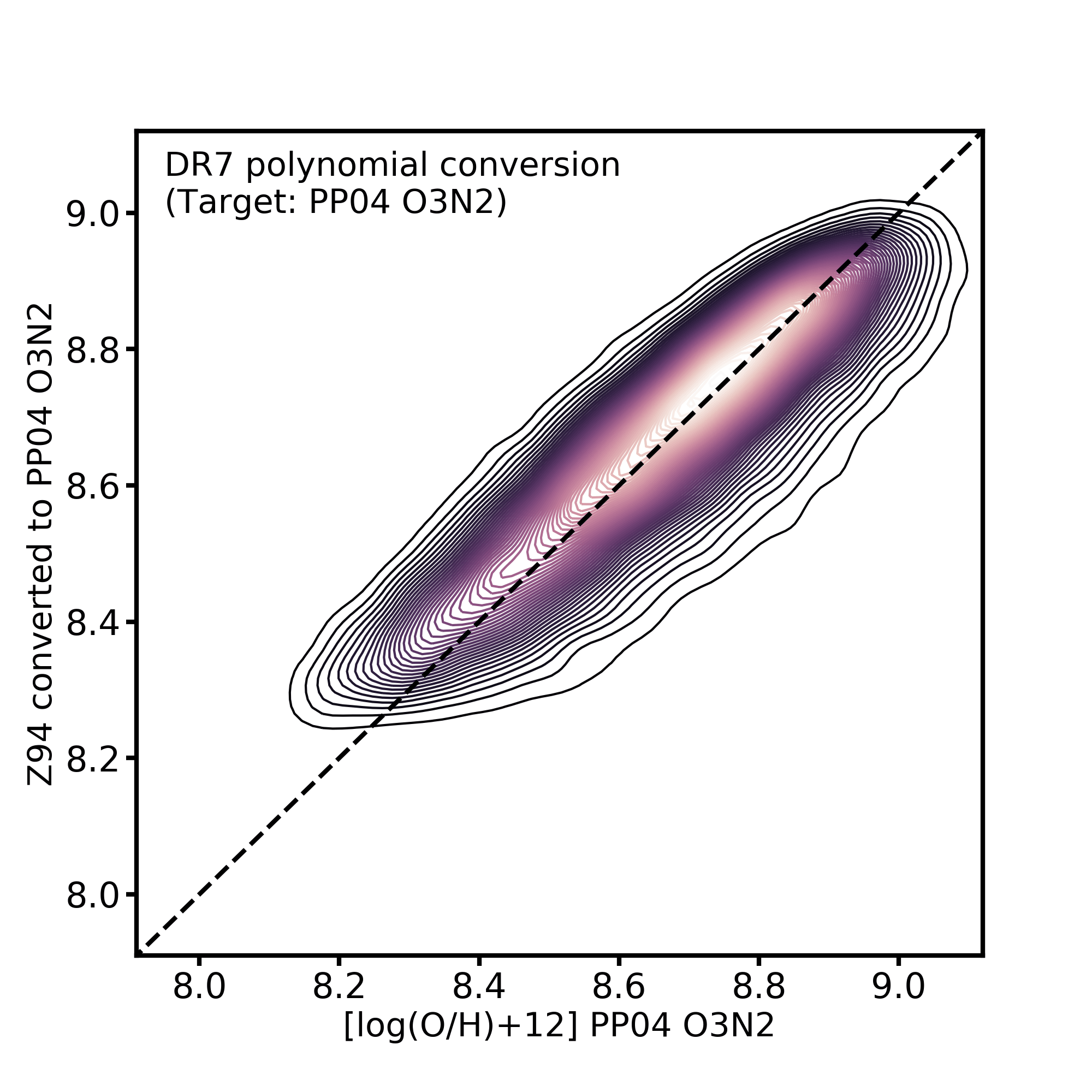}
\includegraphics[width=5.2cm]{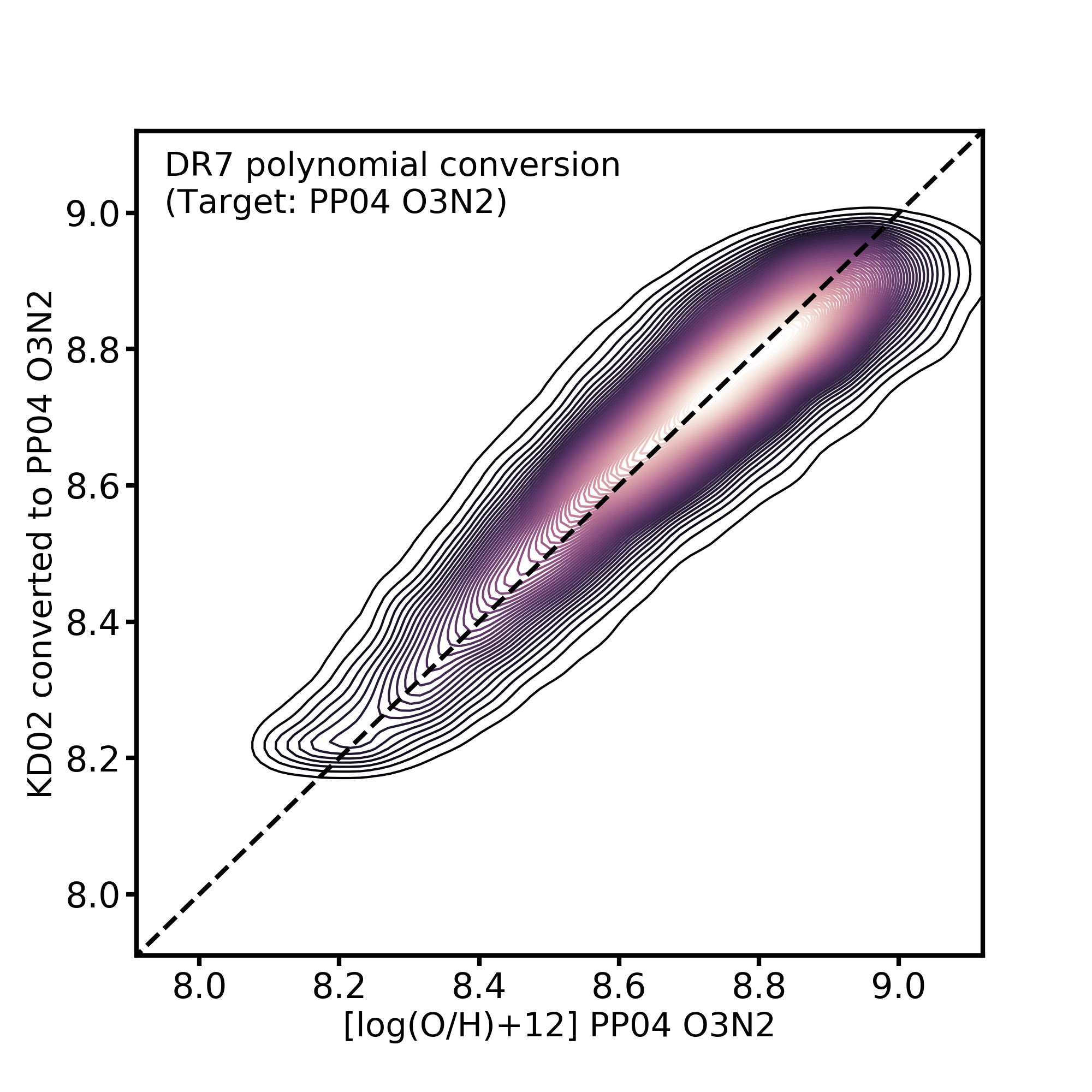}
\includegraphics[width=5.2cm]{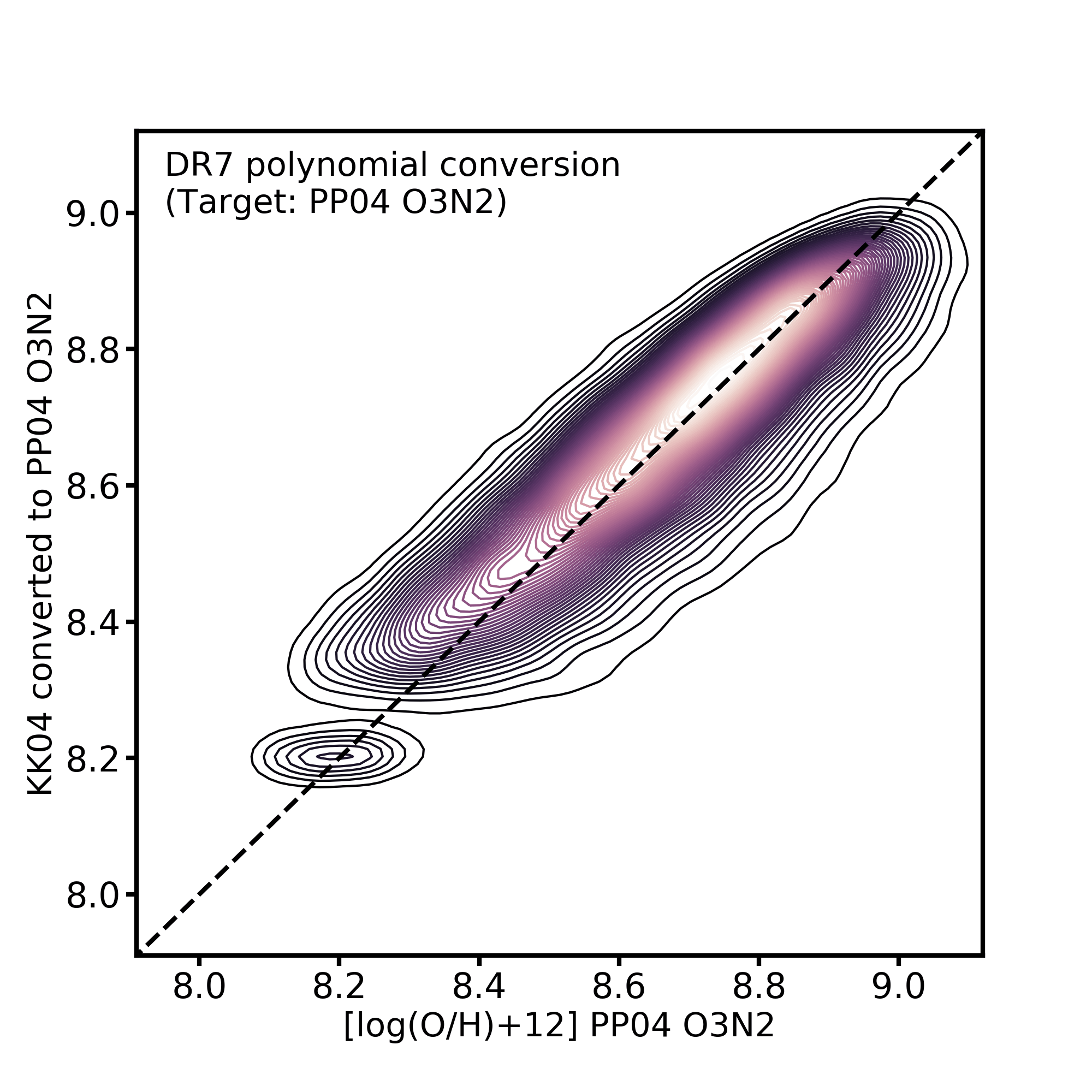}
\includegraphics[width=5.2cm]{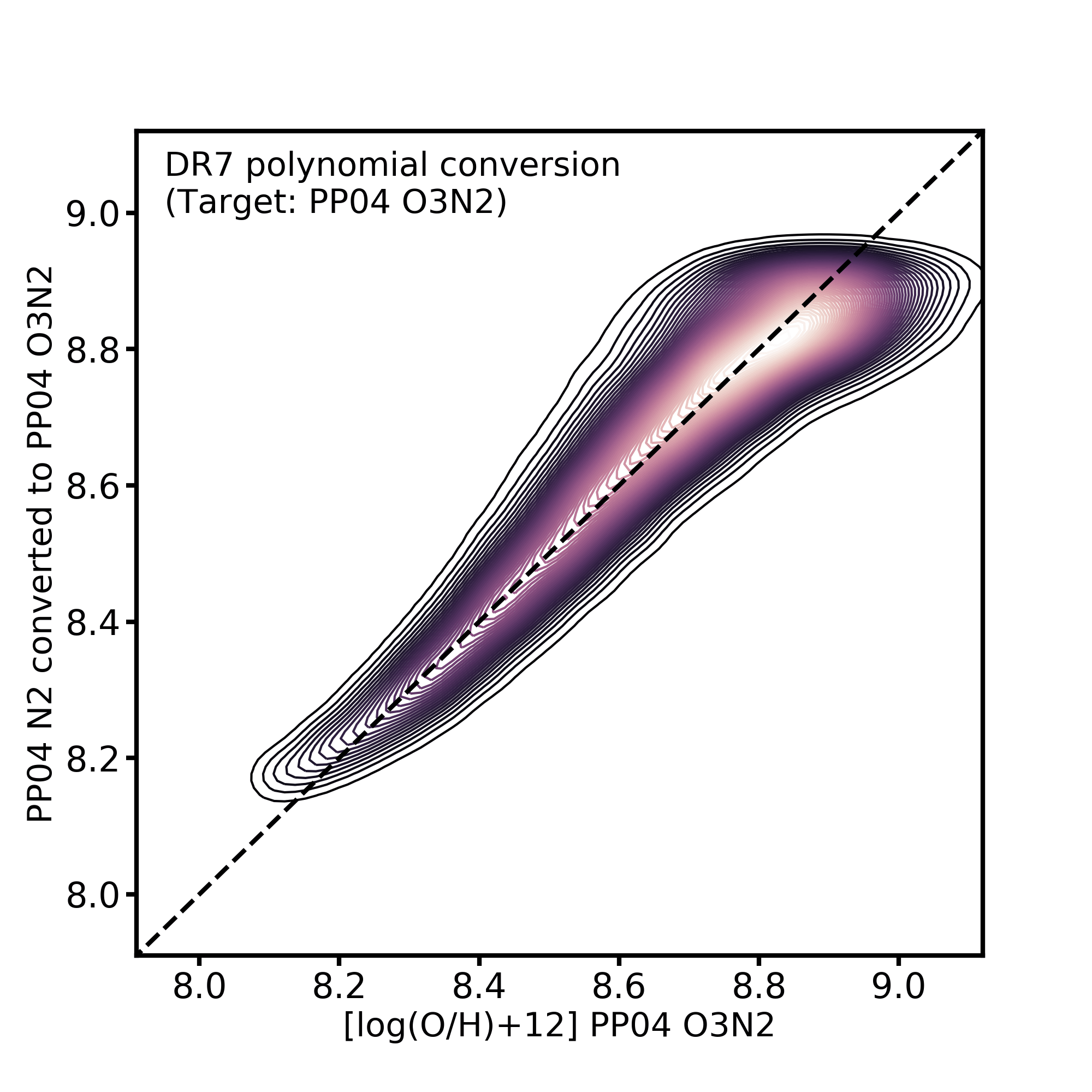}
\includegraphics[width=5.2cm]{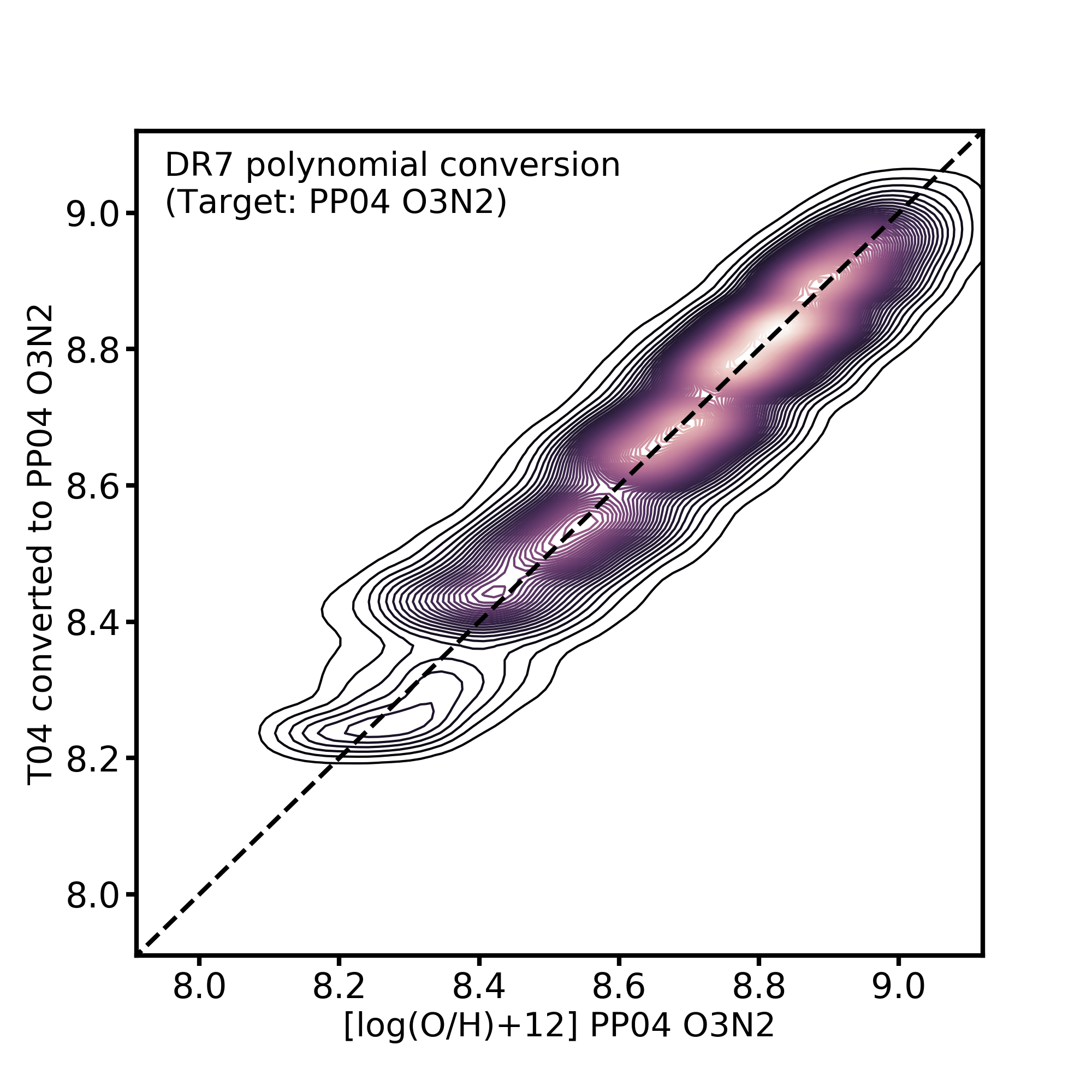}
\caption{The same plot as Fig\ref{fig-old}, but with the newly determined polynomial coefficients derived from the DR7 dataset.  The systematic offsets found in Fig\ref{fig-old} using the DR4 coefficients have now been removed and the converted metallicities perform well.}
\label{fig-new}
\end{figure*}

Derived from a sample of $\sim$ 28,000 star-forming galaxies from the SDSS DR4, \cite{Kewley-08} presented tabulated coefficients for a third order polynomial that could be used to convert between different metallicity calibrations.  We begin by re-assessing the validity of these metallicity conversions as applied to our larger DR7 dataset.  We will then extend these polynomial based conversions to include the five additional diagnostics summarized in Section \ref{sec:new_calibs}.

\subsection{A re-assessment of Kewley \& Ellison (2008) conversions using SDSS DR7 data}
\label{sec:poly_ke08}

Of the ten original metallicity diagnostics explored in \cite{Kewley-08}, conversions between eight of the strong line metallicity calibrations were presented in \cite{Kewley-08}. Of these eight, we have removed Denicolo et al. (2002), for the reasons described in Sec. \ref{sec:ke08_review}, leaving the following seven strong line diagnostics: M91, Z94, KD02, KK04, PP04 N2, PP04 O3N2 and T04.  Gas phase metallicities are computed for the first six of these diagnostics (the T04 metallicity is taken directly from the MPA/JHU catalog) for our DR7 sample of galaxies (see the last column of Table \ref{tab:metal-ref} for the number of galaxies available for each diagnostic).

Fig. \ref{fig-old} shows some examples of the \cite{Kewley-08} conversion coefficients (derived for DR4) as applied to our DR7 dataset.  For this exemplar, we have selected the PP04 O3N2 as the target diagnostic, which is shown on the x-axis of each panel of Fig. \ref{fig-old} for the DR7 dataset.  Each of the other six diagnostics is then converted onto the PP04 O3N2 system using the polynomial coefficients derived by \cite{Kewley-08}.  If the conversions worked perfectly, then the converted version of a given calibration (y-axis) should be identical to its directly measured value (x-axis), and the points will line up along the diagonal 1:1 dashed line.

\medskip

The results shown in Fig. \ref{fig-old} (and a thorough assessment of all the combinations of targets and conversions, not shown here for brevity) confirm that the \cite{Kewley-08} DR4-derived coefficients are not optimized for the DR7 dataset.  The offsets between the converted and target metallicities are typically quite small, $<$ 0.1 dex, but the large dataset shows that these effects are statistically significant, and depending on the calibration in question, can affect a significant fraction of galaxies.  

We determine that the origin of these offsets and biases is likely due to changes in the emission line flux values, which are typically higher in the DR7 than in the DR4. While uniform enhancements in the line fluxes should not produce systematic biases in metallicity values, we find that not all emission lines are stronger in the DR7 by the same percentage, which will present a bias when comparing metallicities based on different emission line ratios. Fig. \ref{fig-old} shows conversions between calibrations which use different emission line ratios, and are therefore strongly affected by this change in line strengths. Conversions between calibrations based on the same emission lines are less discrepant between the two data sets.

We therefore re-compute improved coefficients between the seven metallicity diagnostics in common between our sample and that of \cite{Kewley-08}, using an identical functional form of a third order polynomial.  The new coefficients are tabulated in Tables \ref{tab:M91} -- \ref{tab:T04}.  Fig. \ref{fig-new} shows the same combination of conversions as Fig.\ref{fig-old}, but using these updated polynomial coefficients.  The systematic offsets and biases that were present in Fig.\ref{fig-old} have now been largely eliminated.

\subsection{Polynomial coefficients for converting five additional diagnostics}
\label{sec:poly_new}

\begin{figure*}
\centering
\includegraphics[width=4.2cm]{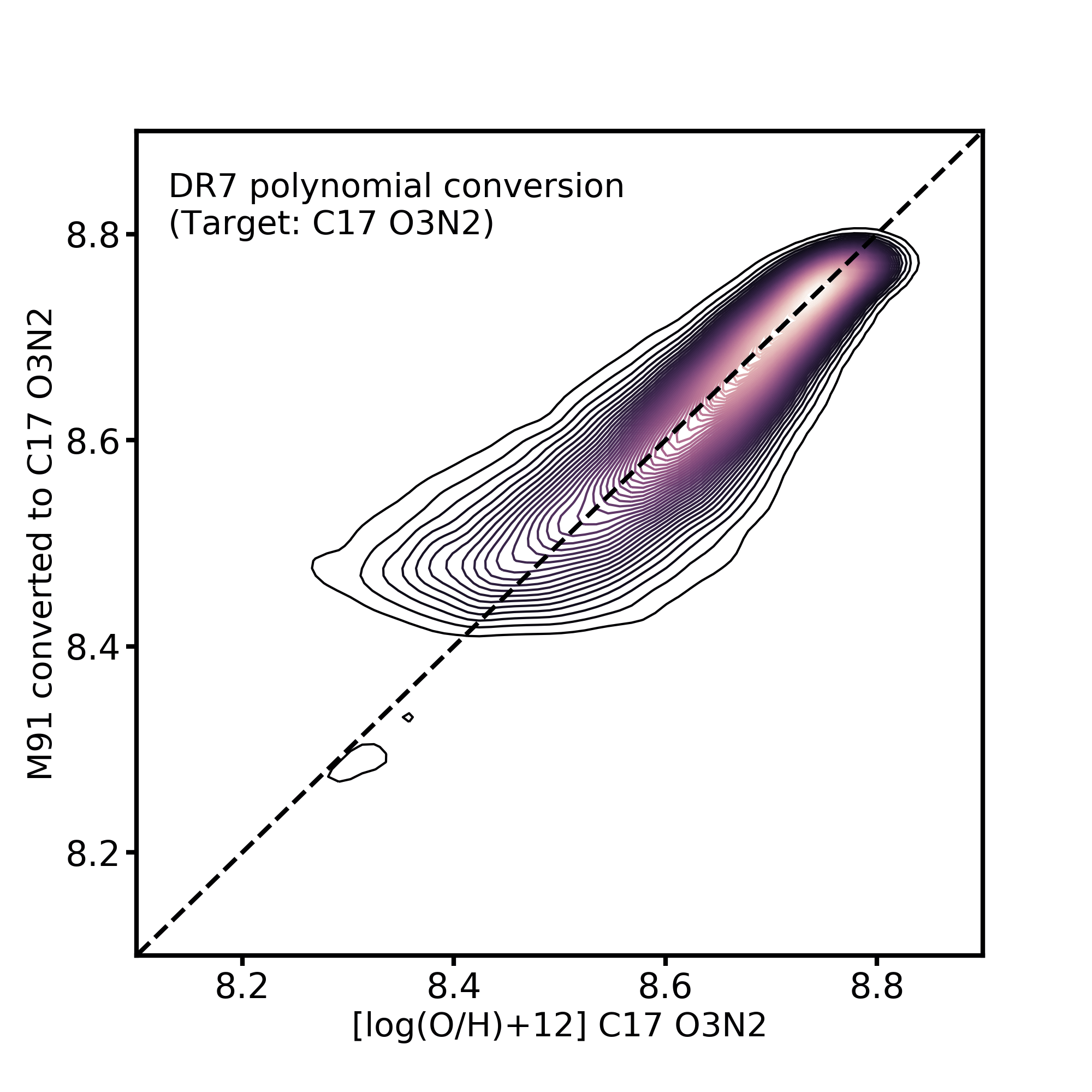}
\includegraphics[width=4.2cm]{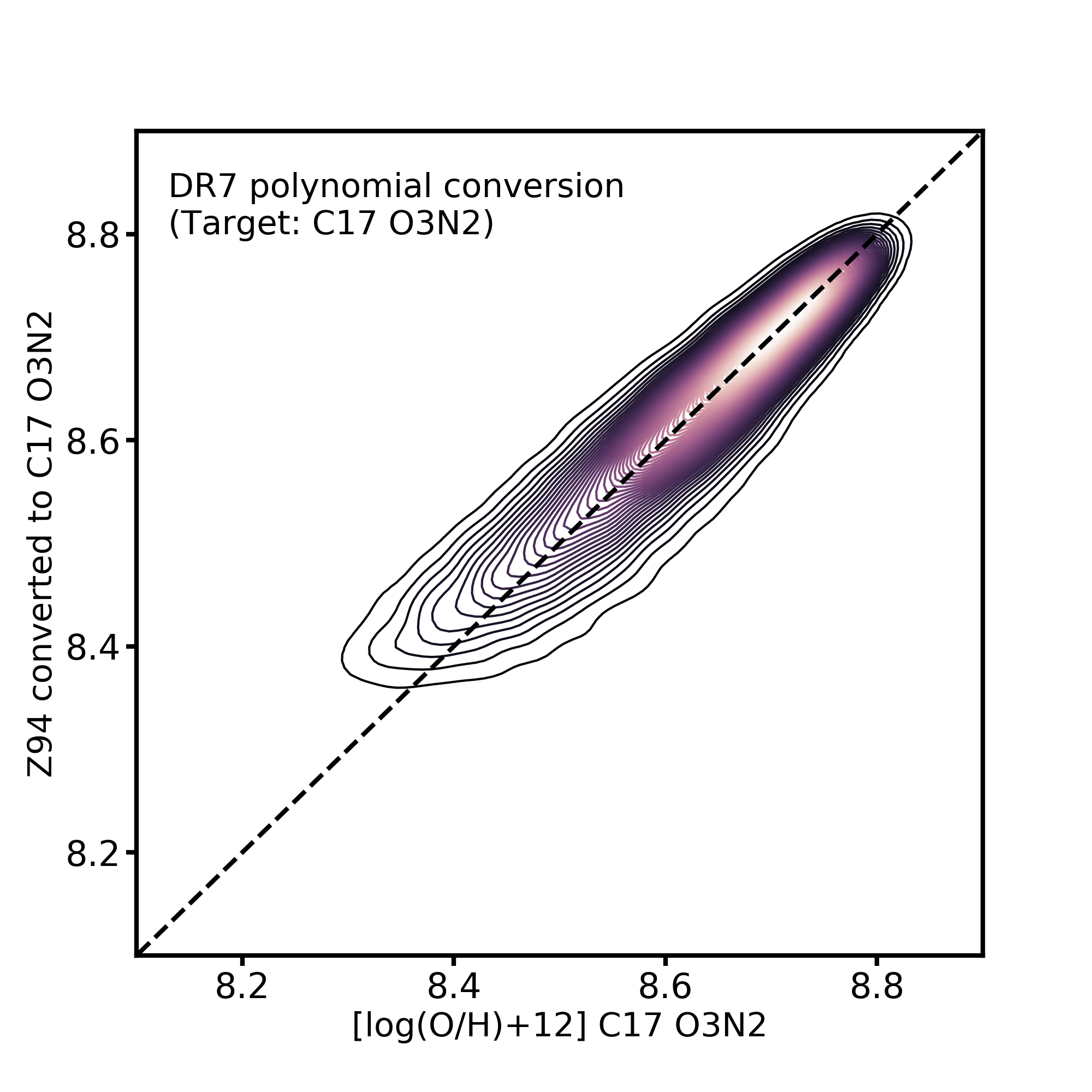}
\includegraphics[width=4.2cm]{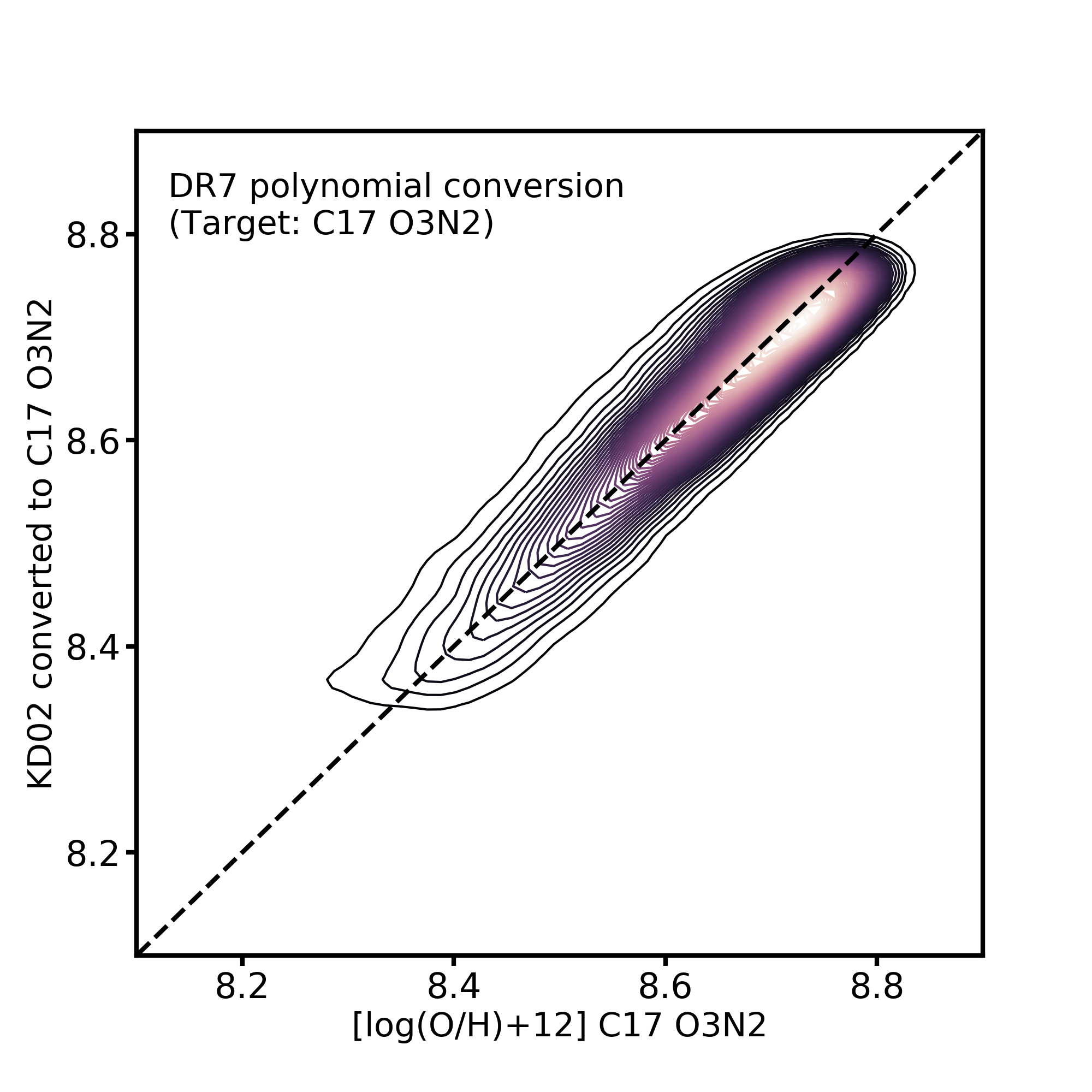}
\includegraphics[width=4.2cm]{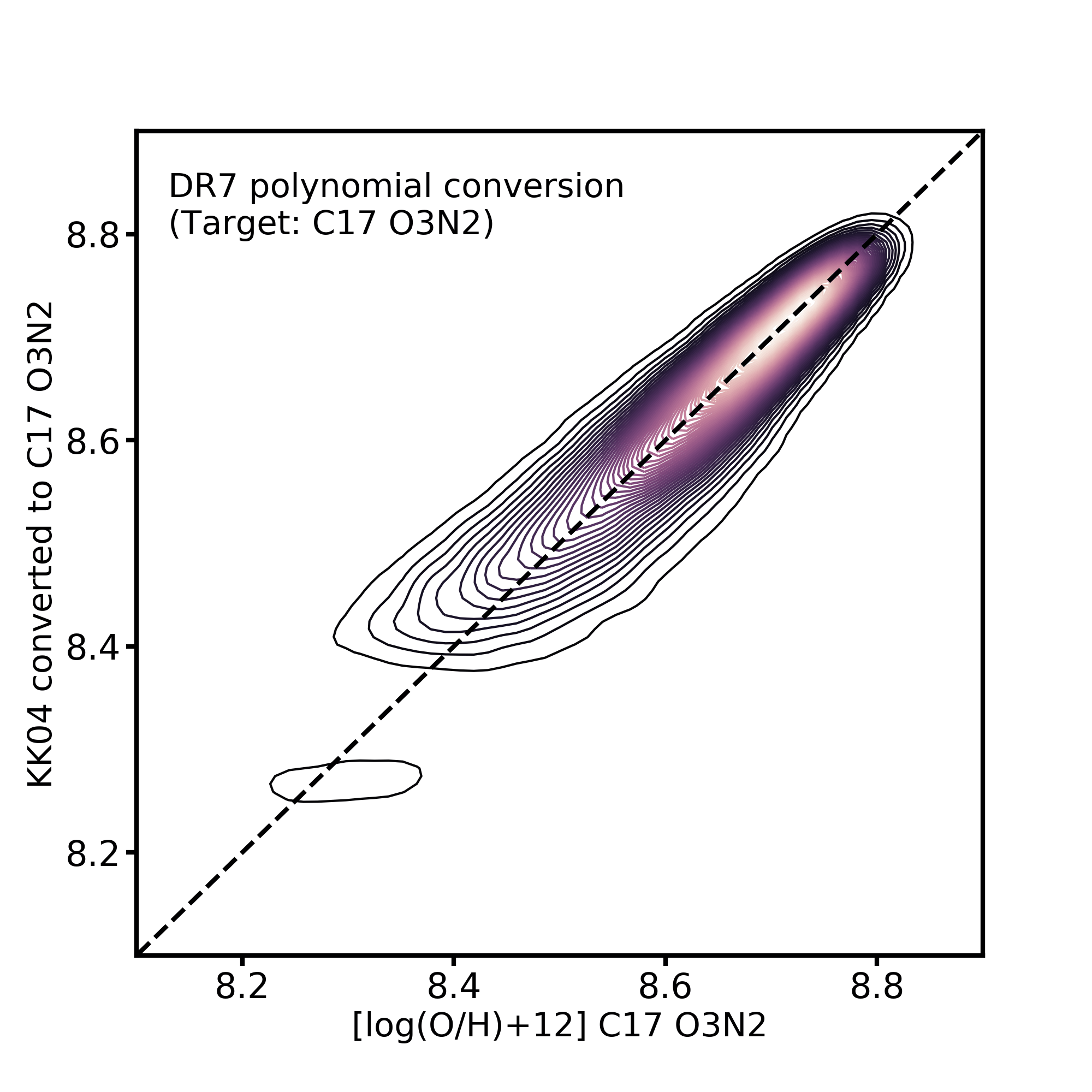}
\includegraphics[width=4.2cm]{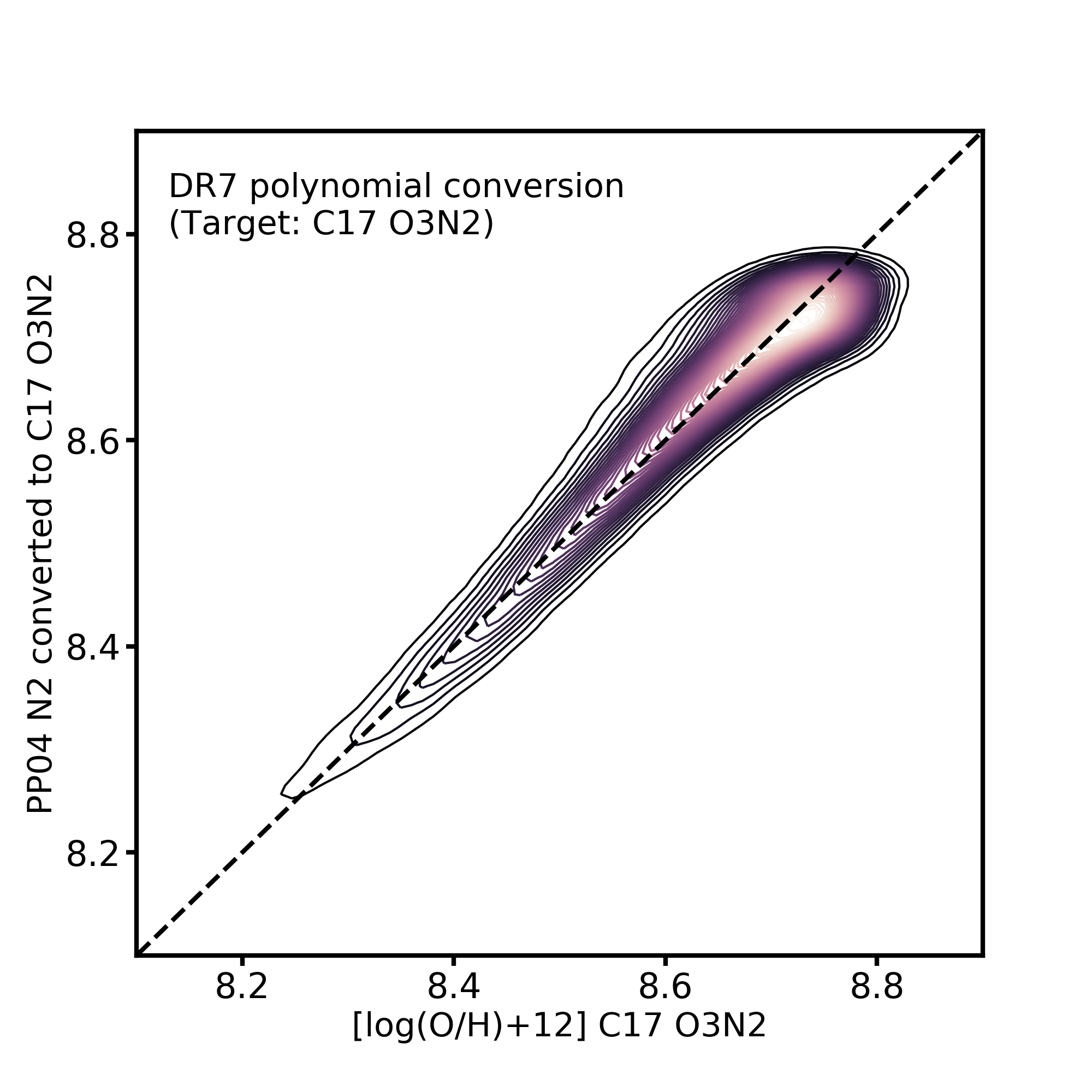}
\includegraphics[width=4.2cm]{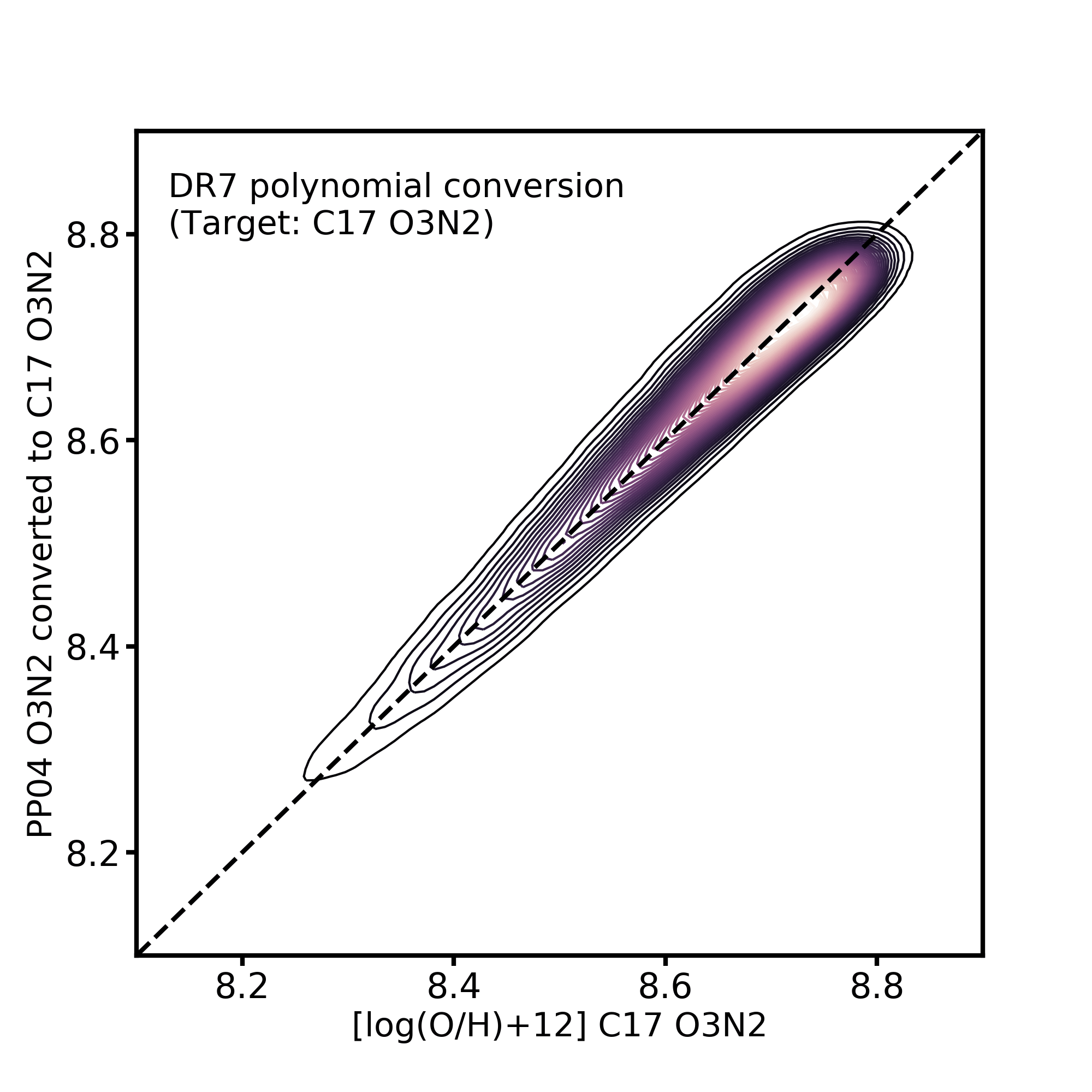}
\includegraphics[width=4.2cm]{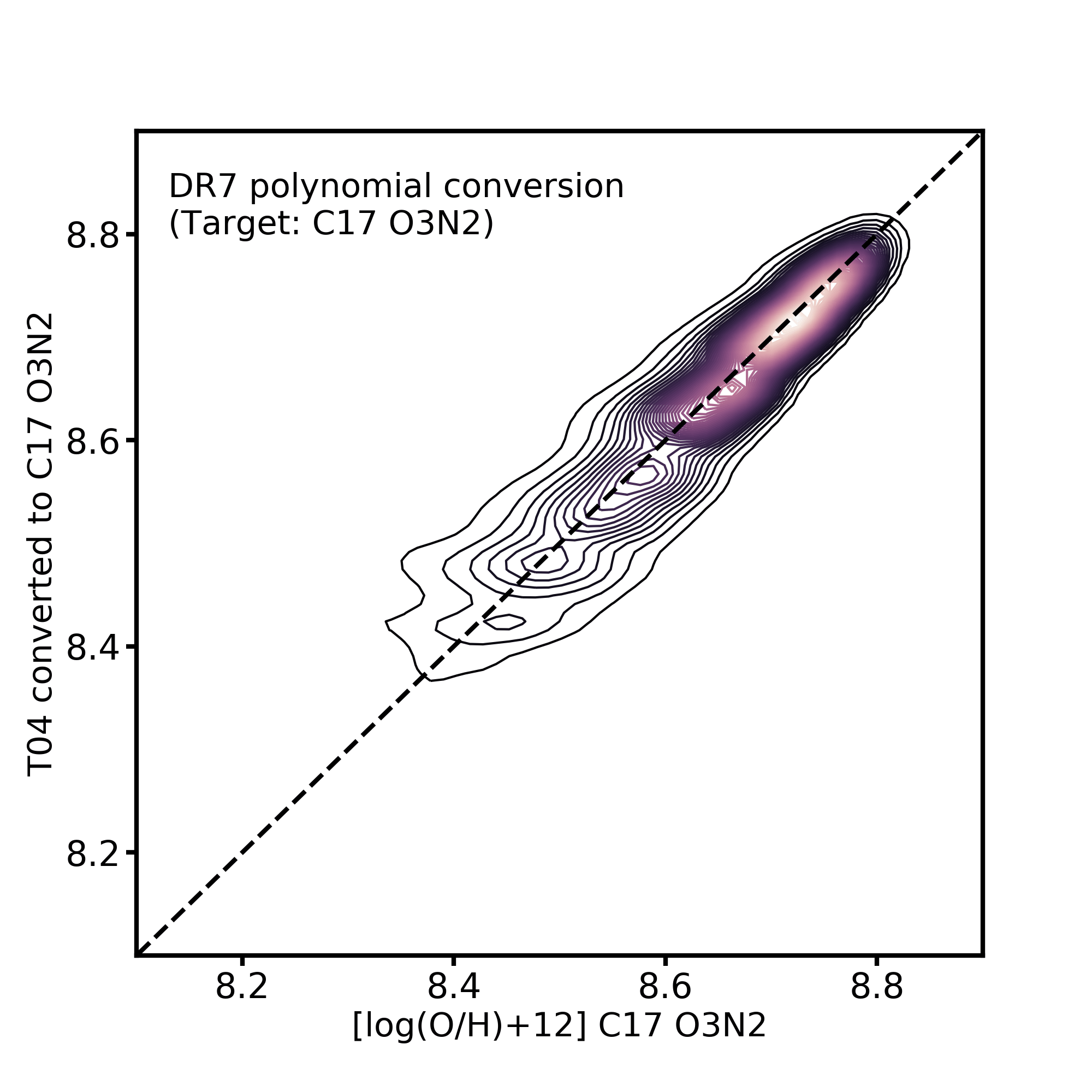}
\includegraphics[width=4.2cm]{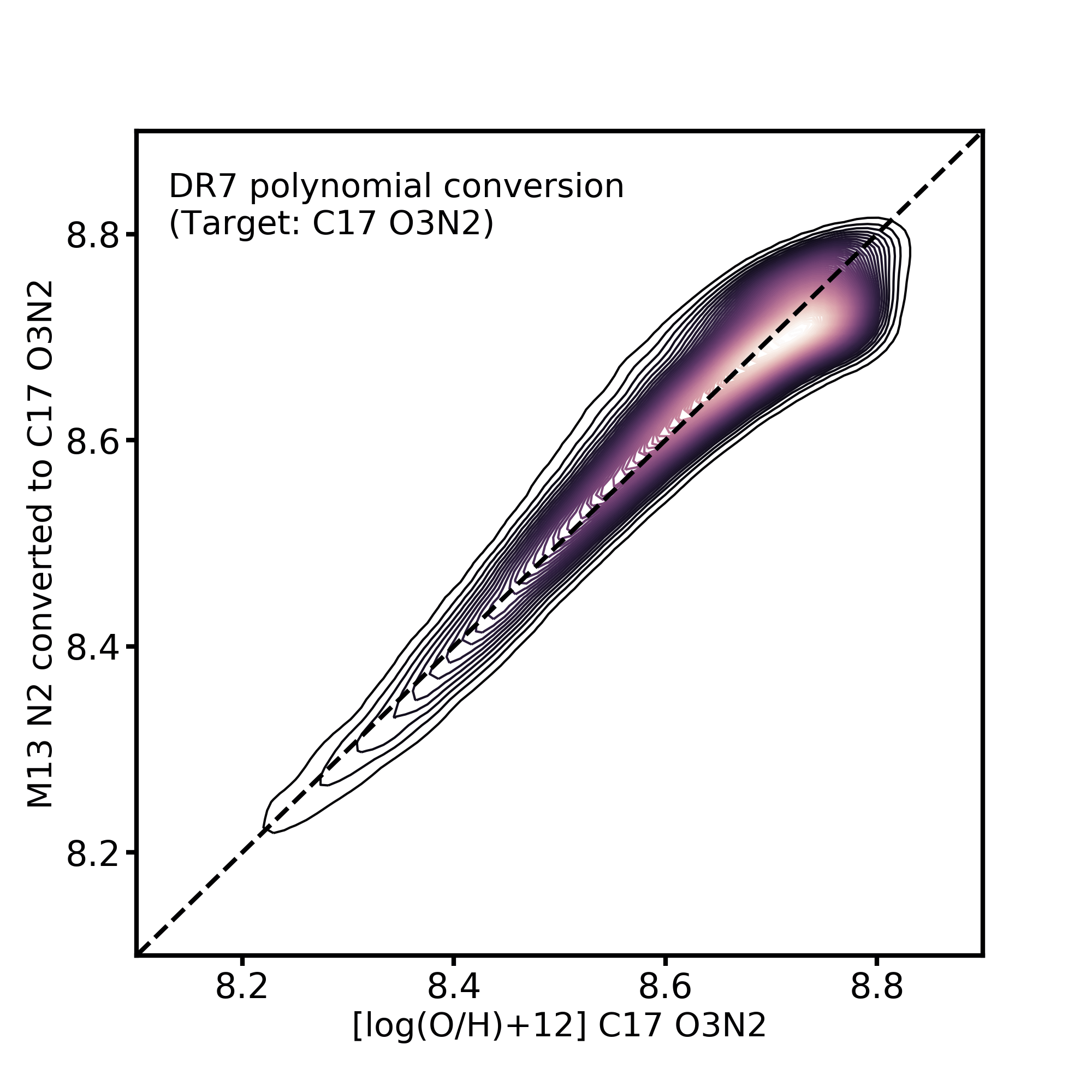}
\includegraphics[width=4.2cm]{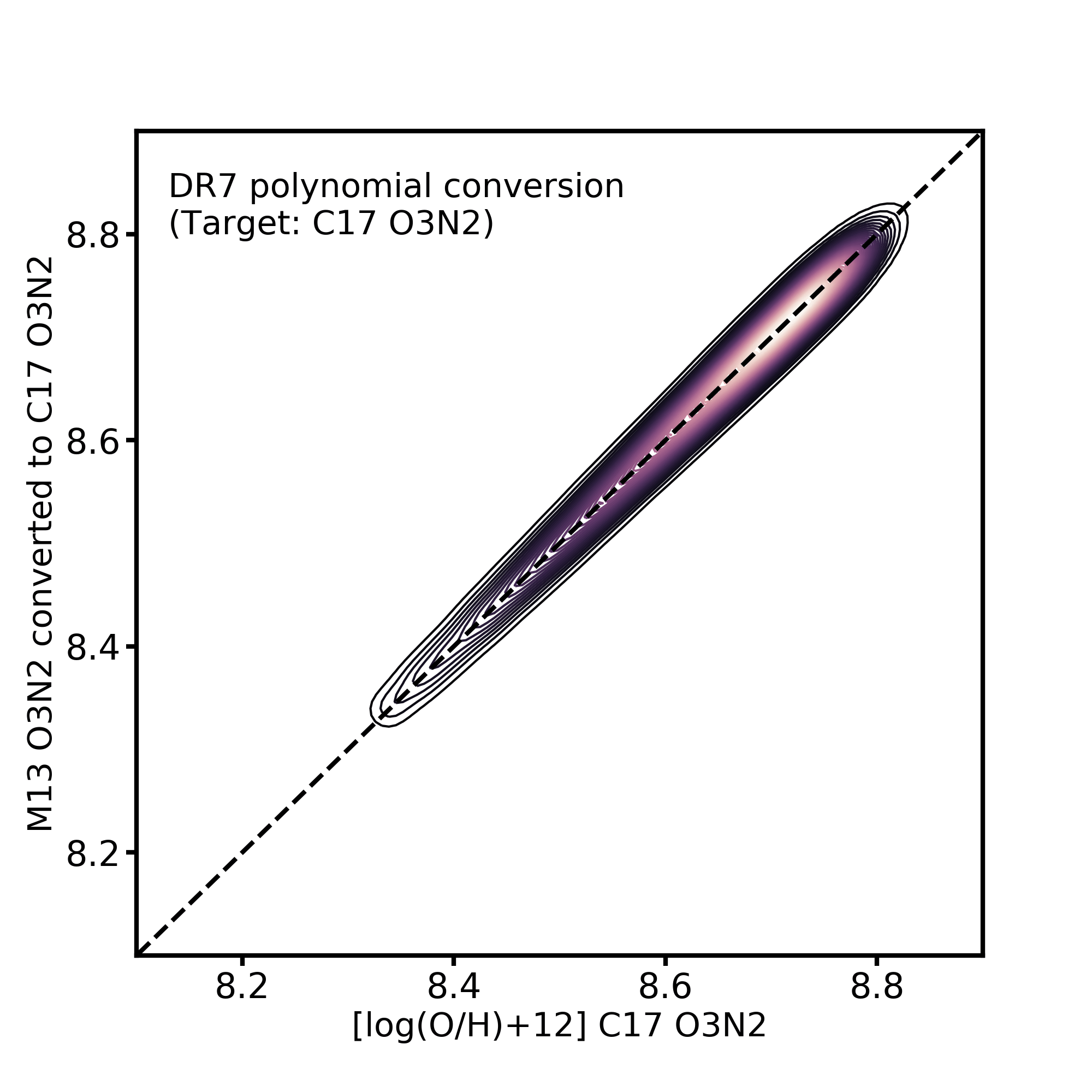}
\includegraphics[width=4.2cm]{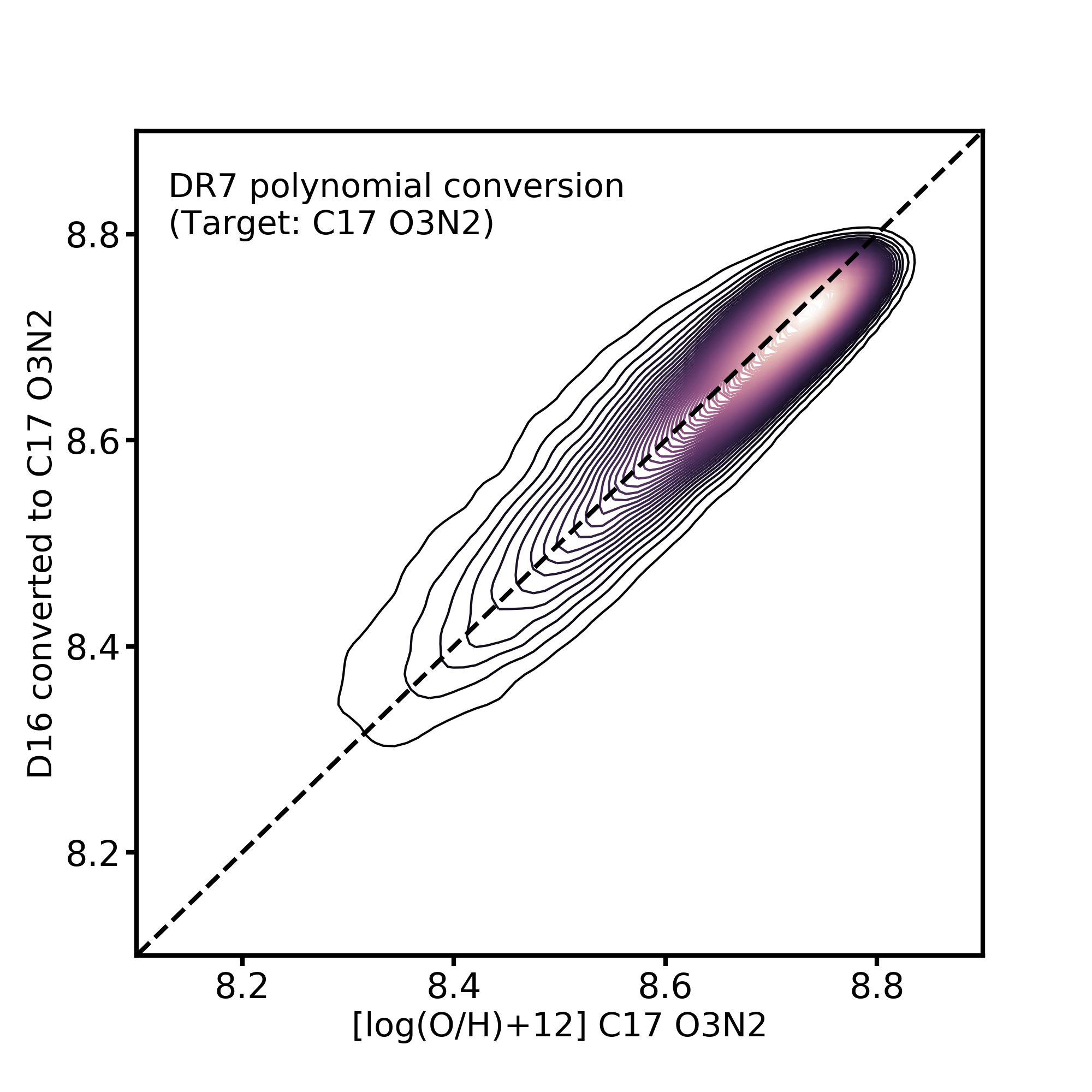}
\includegraphics[width=4.2cm]{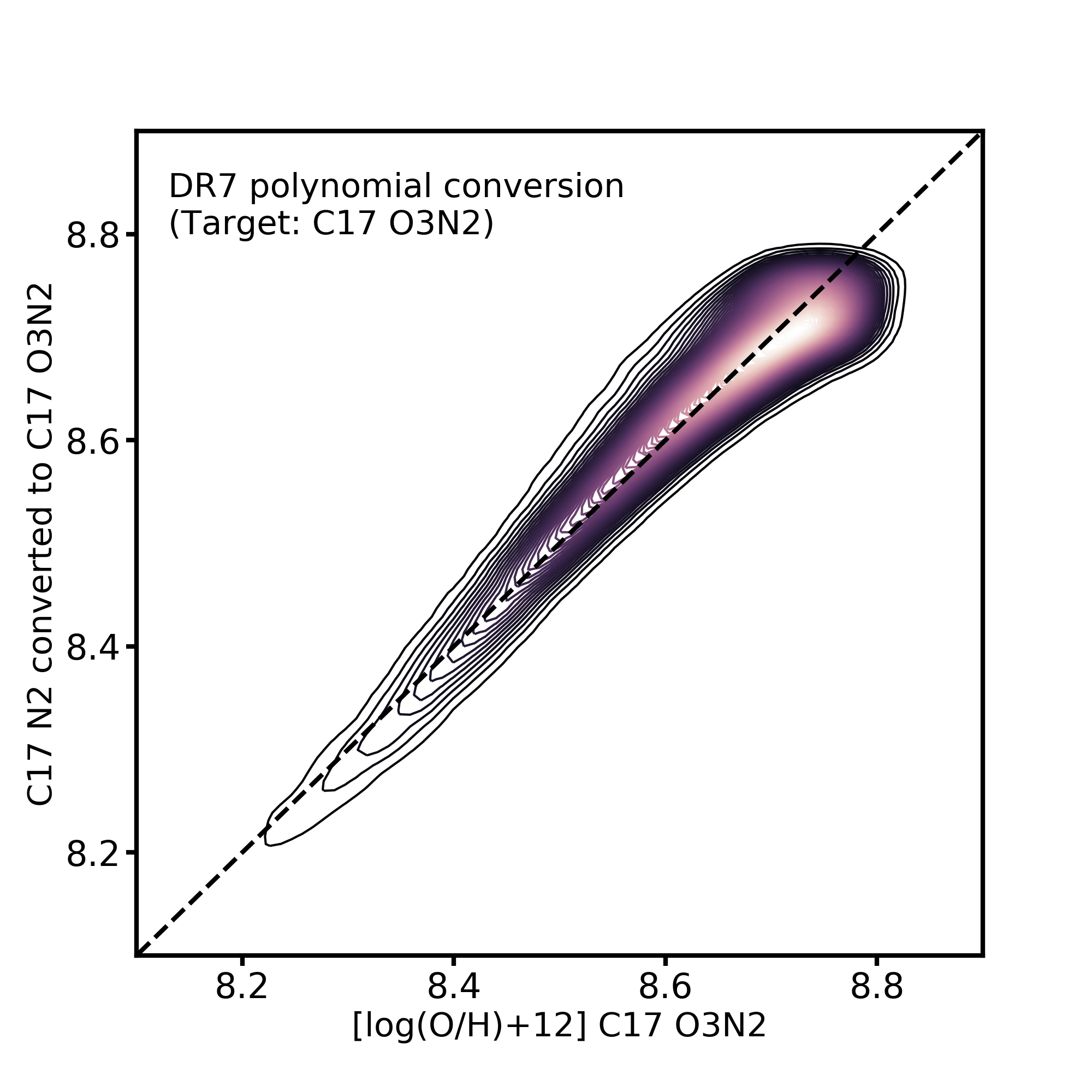}

\caption{Metallicity conversions using the polynomial method and DR7 data for an expanded set of diagnostics.  In this example, C17 O3N2 is the target metallicity diagnostic and the other 11 calibrations have been converted to it using the coefficients in the Appendix tables. }
\label{fig-C17}
\end{figure*}

In addition to the diagnostics previously presented in \cite{Kewley-08}, we have additionally derived conversions, using the polynomial method, for the five additional diagnostics summarized in Sec \ref{sec:new_calibs} (M13 N2, M13 O3N2, D16, C17 N2 and C17 O3N2).  We again use a third-order polynomial, in order to be consistent with the other diagnostic conversions.  The coefficients for these new diagnostic conversions are presented in Tables \ref{tab:M13-N2} - \ref{tab:C17-O3N2}.

In Fig. \ref{fig-C17}, we show an example of these new conversions, using C17 O3N2 as the target metallicity diagnostic.  Similar to Figs \ref{fig-old} and \ref{fig-new}, the C17 O3N2 metallicities (derived directly from the SDSS DR7 spectra) in a given calibration are plotted on the x-axis, with each of the other diagnostics, converted into C17 O3N2 via their polynomial fit, are plotted on the y-axis.  The new conversions show no systematic offsets or skewed behaviour, confirming that the third-order polynomial fit between pairs of calibrations is a good representation of the data.

Taken together, the updated coefficients in Tables \ref{tab:M91} -- \ref{tab:C17-O3N2} represent an improvement in both accuracy and available diagnostics over the original work on this topic by \cite{Kewley-08}.  Given the cumbersome nature of the table format, we have also developed a Graphical User Interface (GUI) into which these coefficients are coded, to facilitate the use of the polynomial functions within the community. We note for transparency that the metallicities output by this GUI are rounded to three decimal places. The GUI can be downloaded here: \href{https://drive.google.com/file/d/1Judh3GW3emjLOE6fk-sqken9yOfacDA7/view?usp=sharing}{Polynomial-link}.

\section{The Random Forest method}
\label{sec:RF}

Figures \ref{fig-new} and  \ref{fig-C17} show that, for the 12 strong line metallicity calibrations presented here, a third-order polynomial can be used to convert between diagnostics with no systematic offsets.  None the less, there are several reasons to motivate an investigation of non-linear solutions to these conversions. First, since a polynomial function is a linear model based on coefficients, it is less able to capture any nonlinear features in the correlations between calibrations than a random forest model which is nonlinear by construction.  A further potential advantage of a machine learning approach is that degeneracy breaking decisions (such as upper and lower branch of $R_{23}$) are naturally included in the model (see also the discussion in Ucci et al. 2018; Ho et al. 2019).

\subsection{Methodology}

\begin{figure}
\centering
\includegraphics[width=8cm]{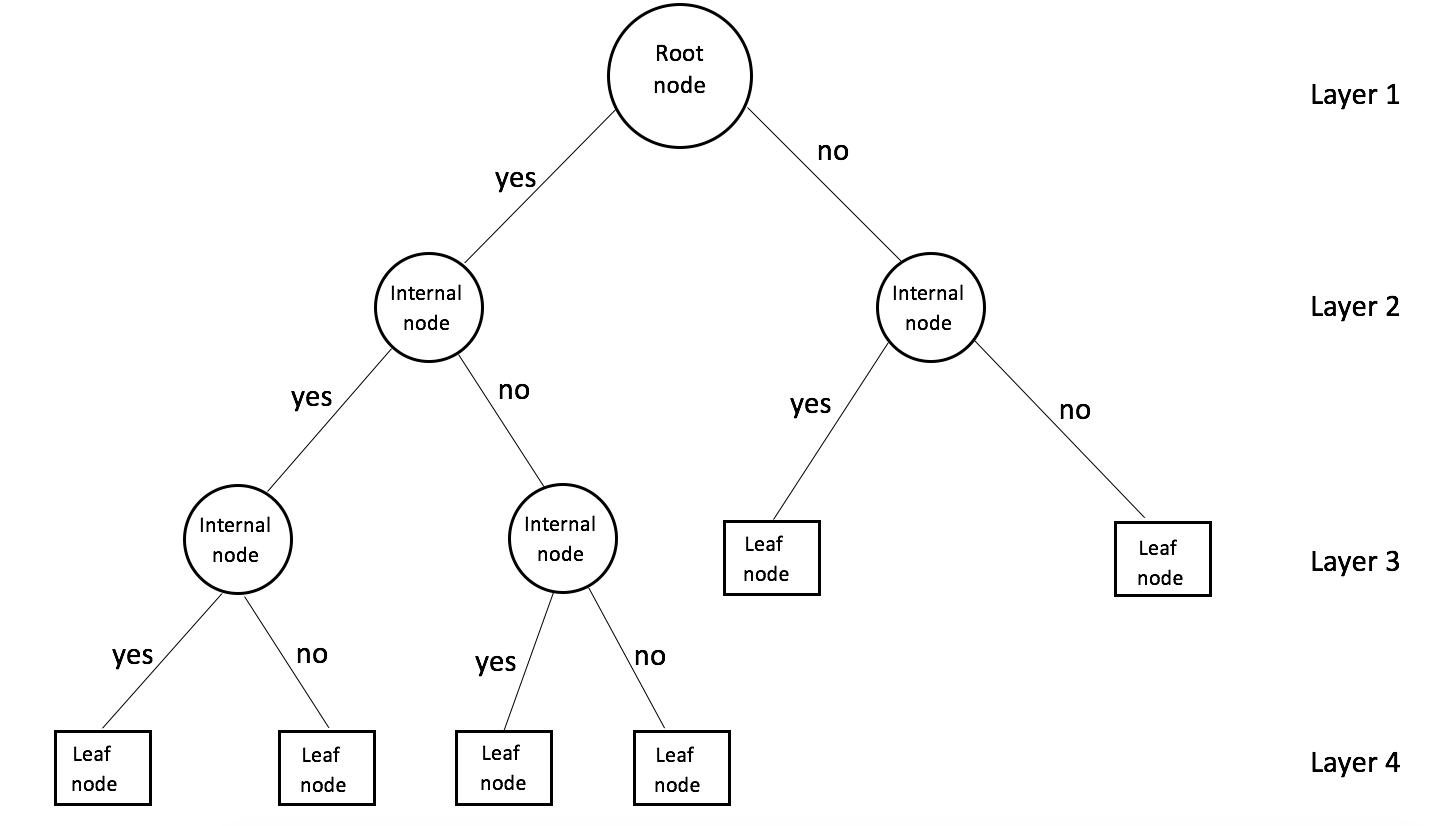}
\caption{A schematic diagram of the structure of a decision tree.  A random forest is comprised of many individual decision trees, which mitigates the problem of over-fitting the data.}
\label{fig-tree}
\end{figure}

A random forest (RF) is a supervised and non-parametric model used for regression and classification problems. In a non-parametric model, there is no internal parameter to train and learn. Since an RF is a supervised method, it implies that for a vector of input parameters, there is a target vector that we want to predict.  An RF consists of several similar decision trees - the building blocks of the random forest model. A decision tree is a system of yes/no options (see Fig.\ref{fig-tree} for a schematic of the process). This system is a recursive partitioning model in which, from the root node, the input data is repeatedly divided and sub-divided.  First, the root node is split into two (internal) nodes based on the binary decisions on an input value of the data under study. Then the procedure is repeated for the two nodes (i.e., in layer-2). In this way, the depth can be increased.  The depth of a tree is a parameter that is chosen by the user. We can stop the tree from growing according to some criteria such as the maximum depth or the minimum samples in a node. The end result is a decision tree with final nodes that are called leaves. Each leaf consists of a set of arranged input parameters and the corresponding target value. In a regression problem, the average value of the targets in a leaf will be predicted in that leaf.  A single tree has the potential for overfitting as we increase the depth of the tree. To tackle this problem, we can extend the use of a single decision tree to a `forest' \citep{Ho-T-98}.

In a RF model, a set (e.g., 100) of similar trees  are made (i.e., with the same depth and the same decision procedure). Input to the first tree is a portion (e.g., 20\%) of the primary input data, which is randomly selected (with replacement) and then fed to the tree. This procedure is repeated for all trees one by one. In this way, each tree's prediction will be a slightly different from other trees due to the randomization procedure. The final regression prediction is the average of all predicted results of all similar trees. This method prevents overfitting and generally gives better and more accurate results.  One advantage of using an RF model is that for a two-branch problem, such as $R_{23}$ metallicity conversions in \cite{Kewley-08}, one single RF model can predict the metallicity values in  branches, without the need for a user-defined break point between the branches.  An RF approach therefore not only reduces the number of required models, but also removes the need for subjective boundaries.  In this work, we use the RF package from scikit-learn \citep{sklearn}.  We use an RF with 100 estimators (trees) and increase the depth of the trees to have at least 10 members (metallicities) in leaves, which is found to be the optimal number for the present work.

\subsection{Results from the random forest}

\begin{figure*}
\centering
\includegraphics[width=4.2cm]{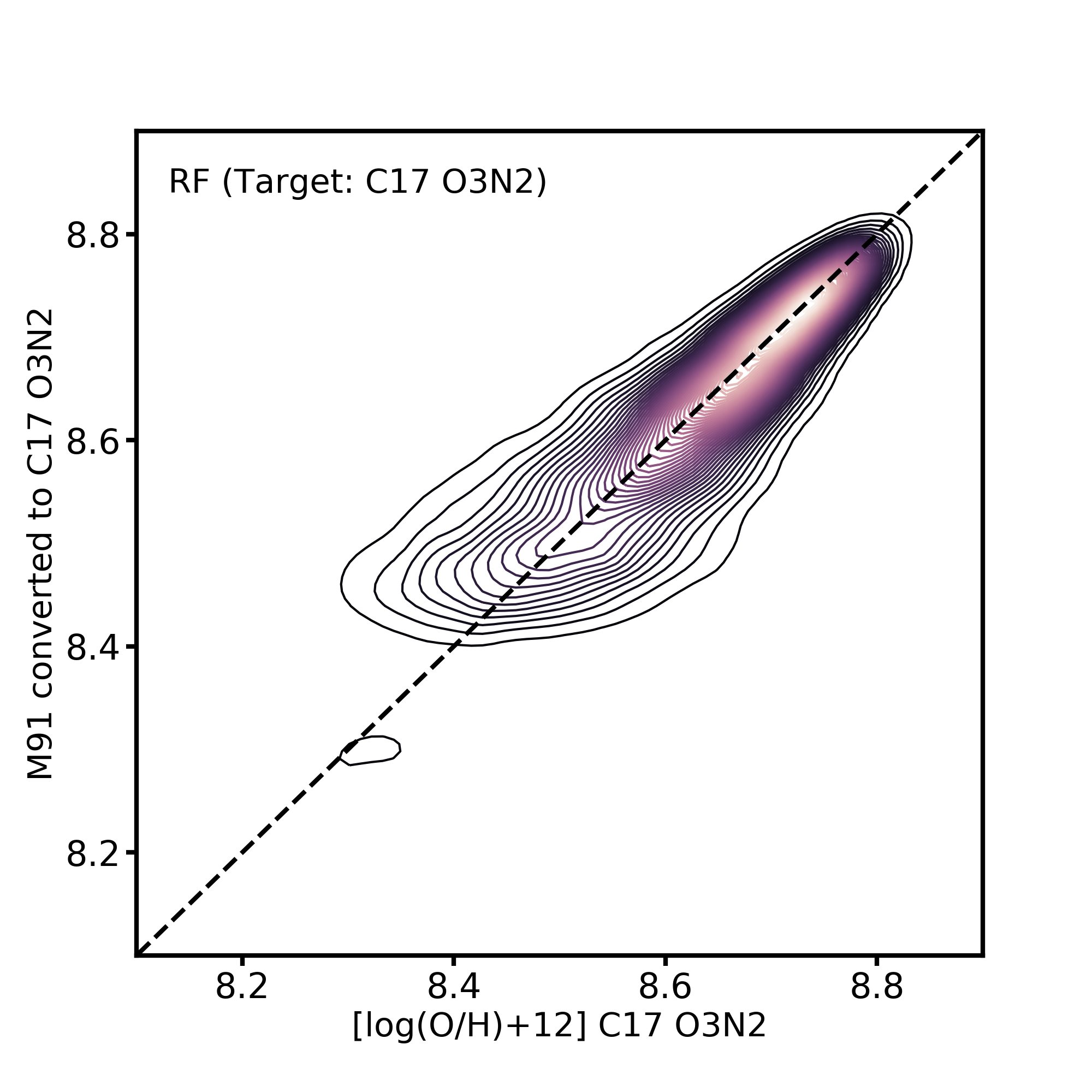}
\includegraphics[width=4.2cm]{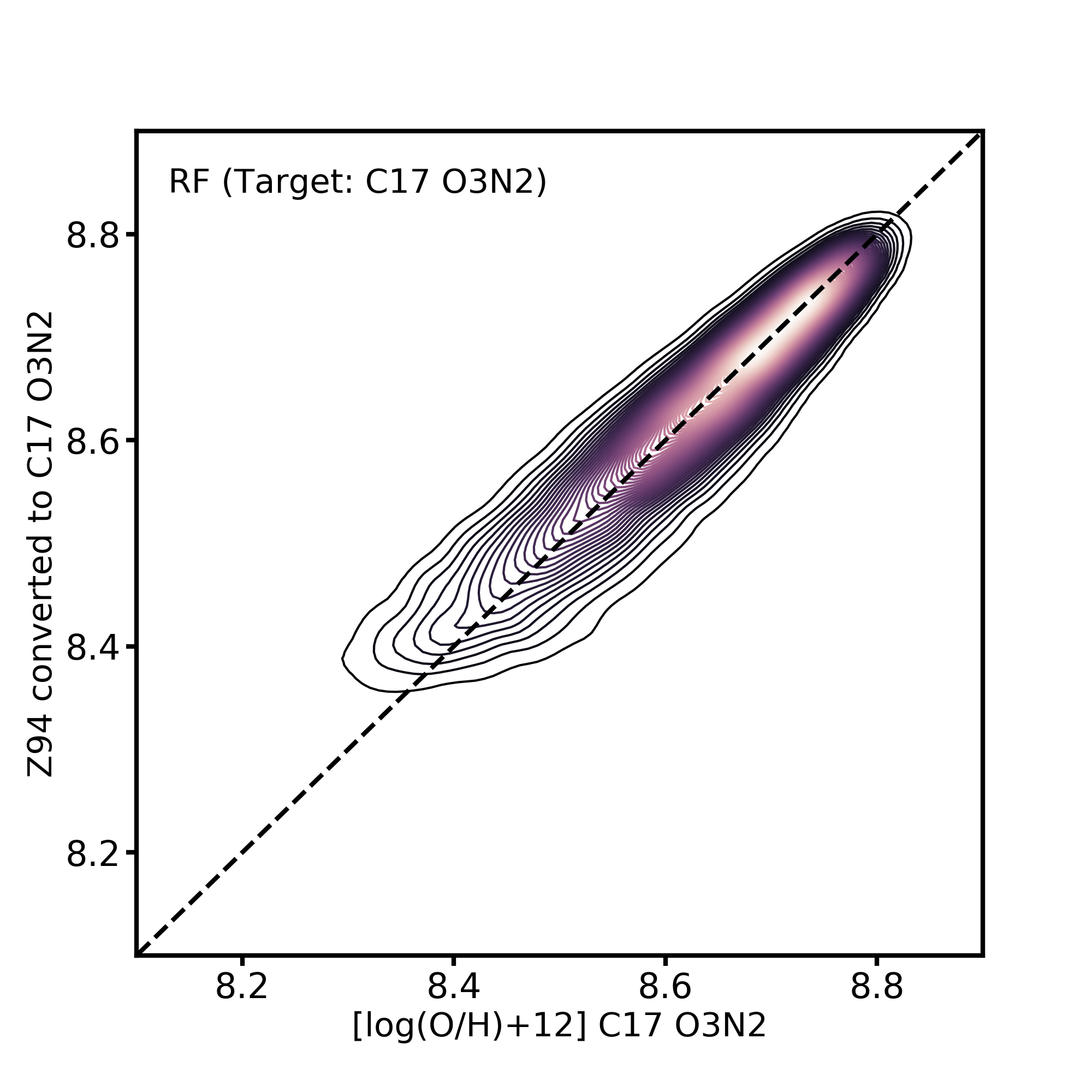}
\includegraphics[width=4.2cm]{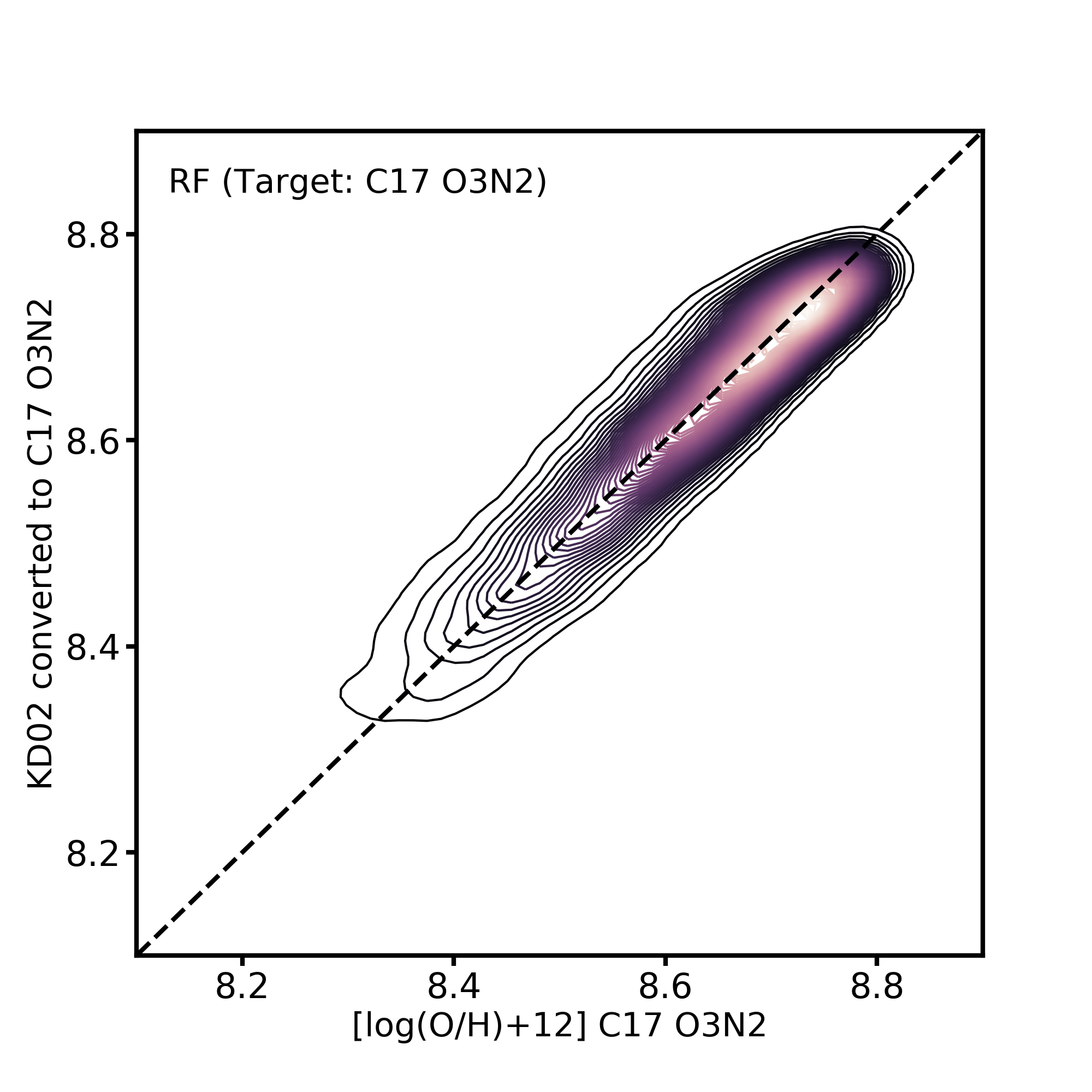}
\includegraphics[width=4.2cm]{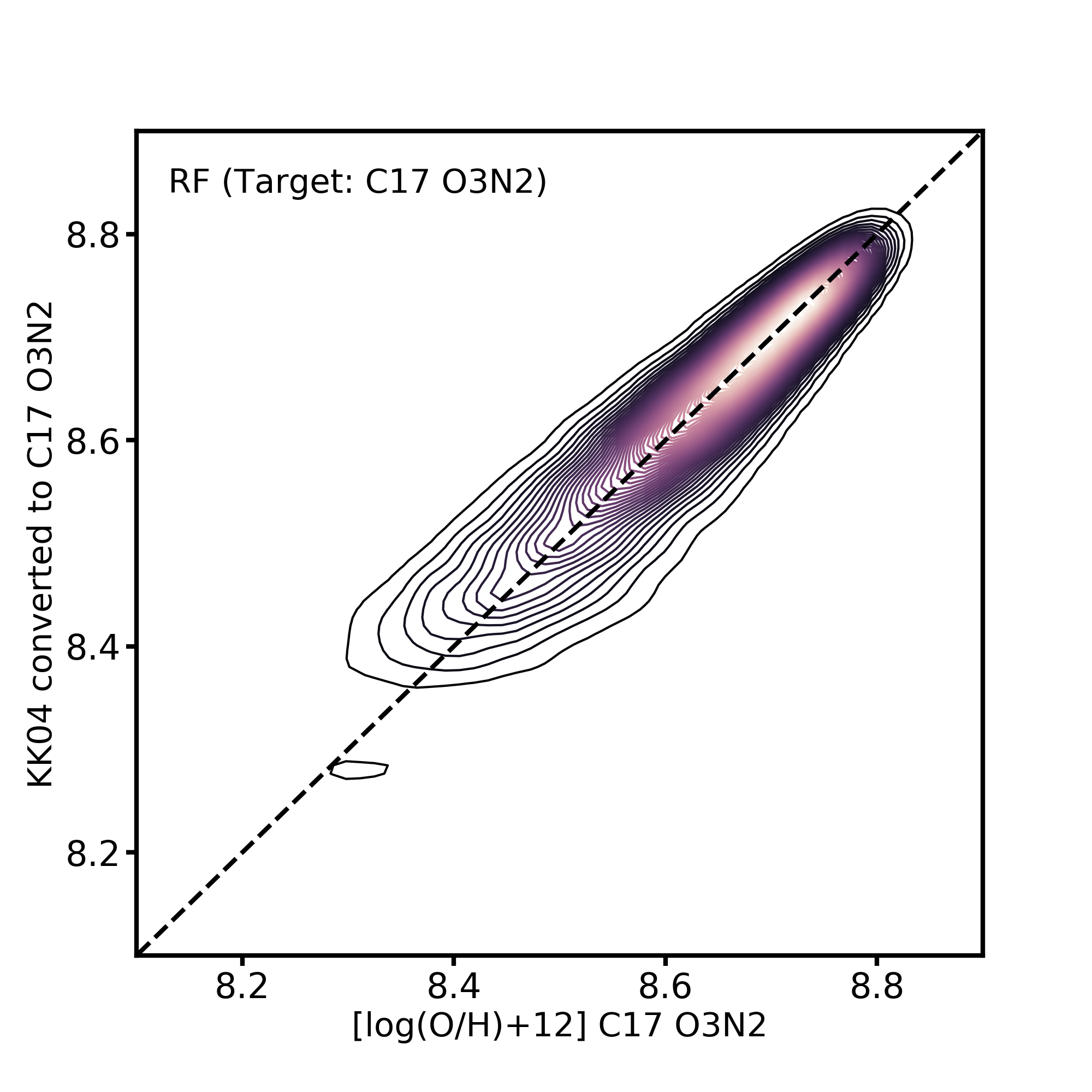}
\includegraphics[width=4.2cm]{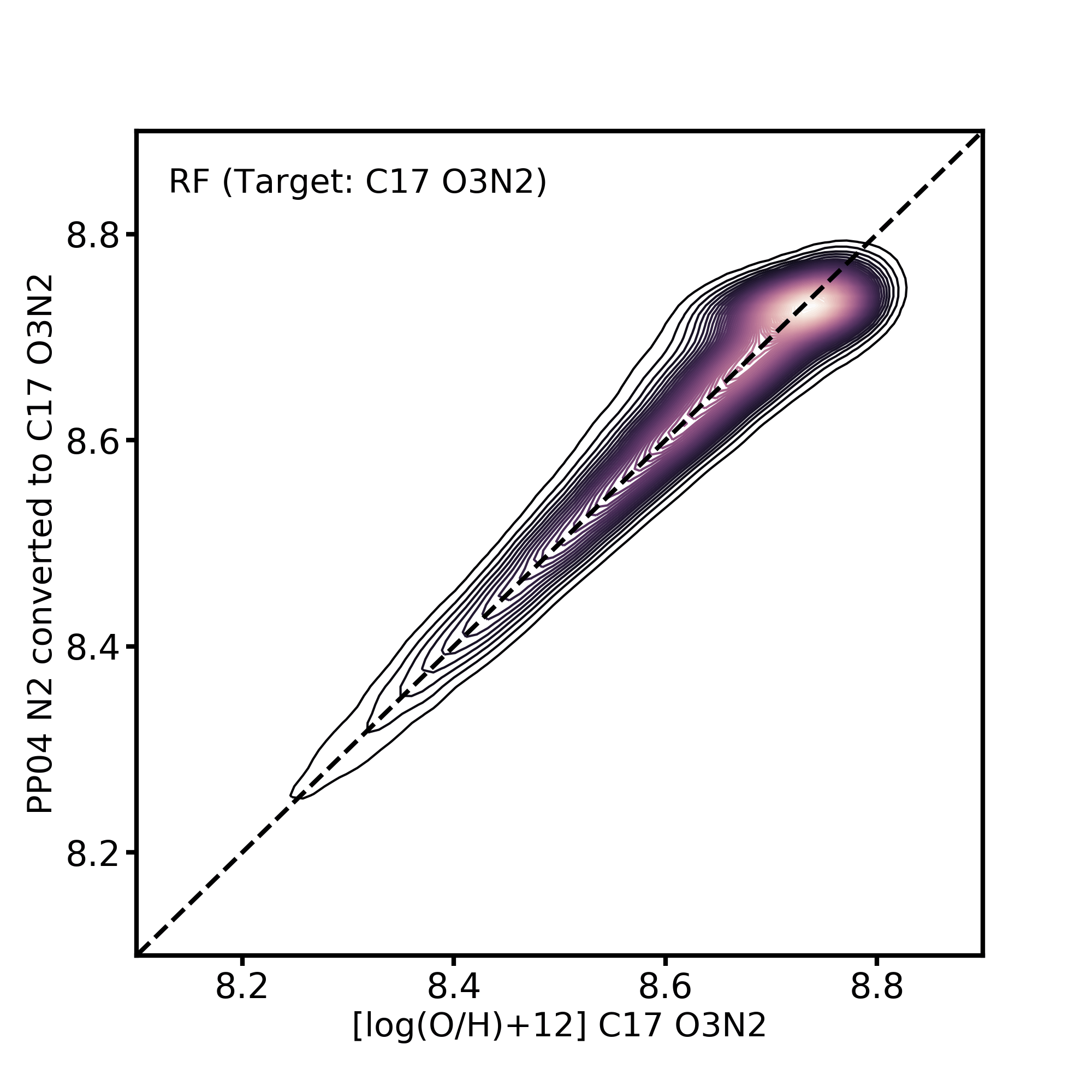}
\includegraphics[width=4.2cm]{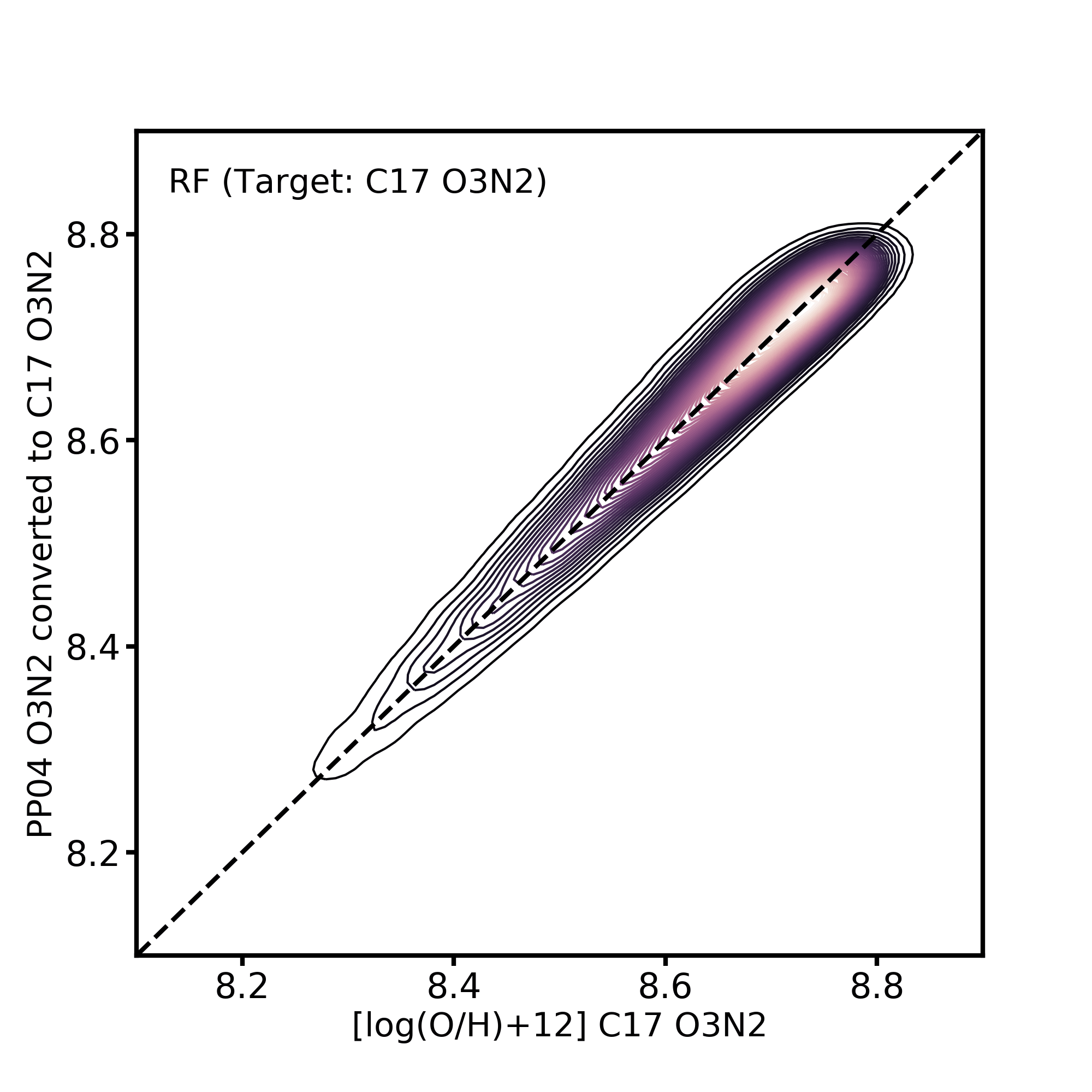}
\includegraphics[width=4.2cm]{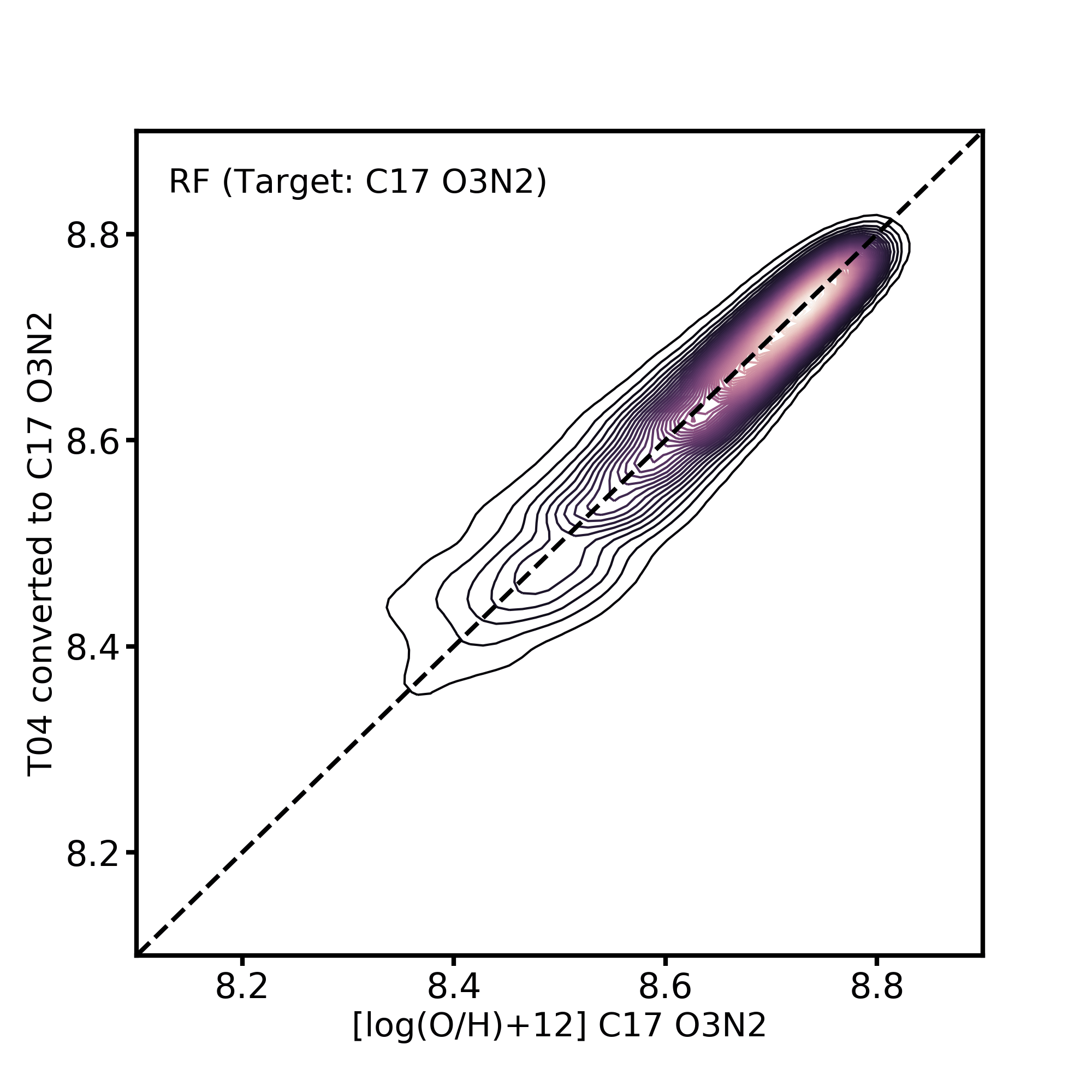}
\includegraphics[width=4.2cm]{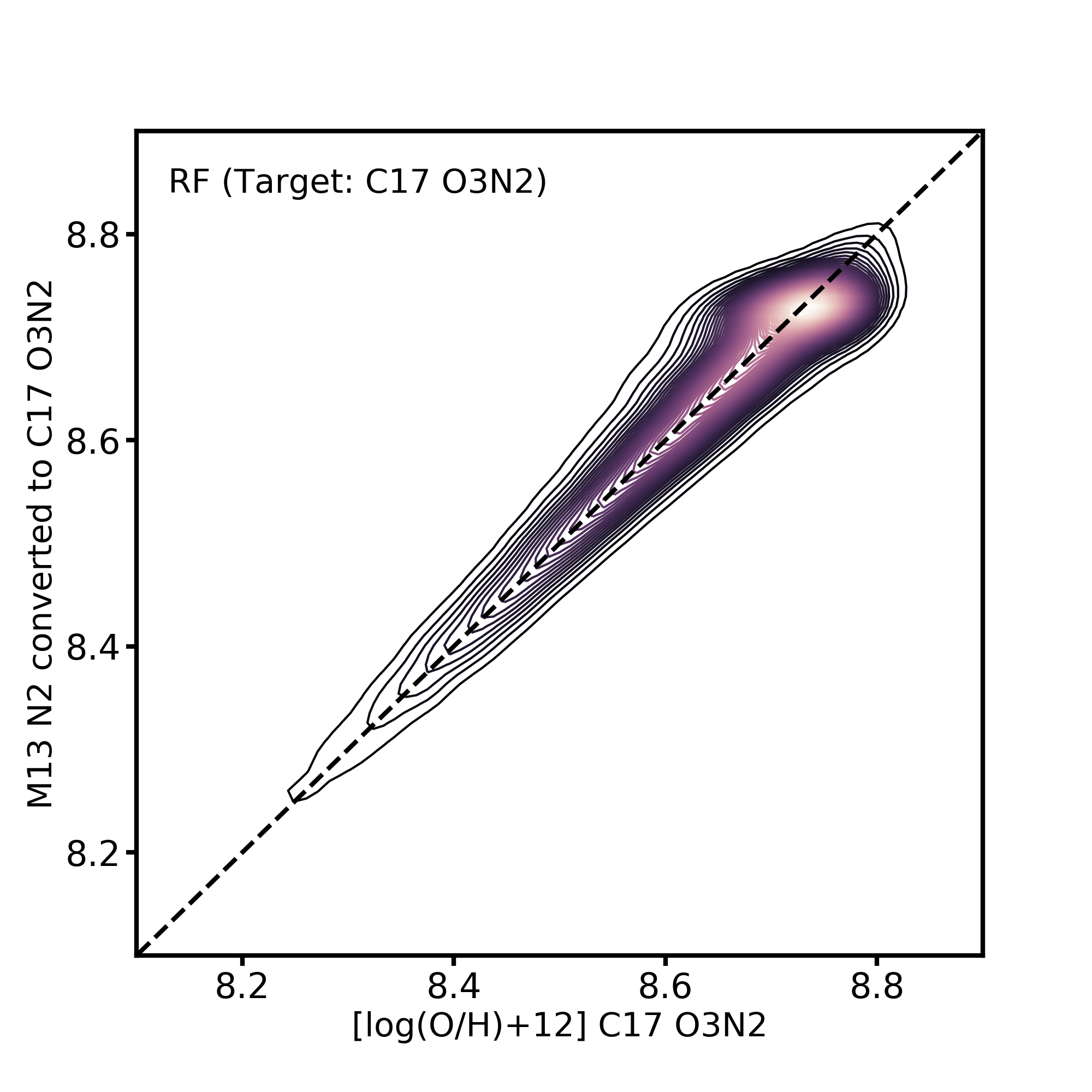}
\includegraphics[width=4.2cm]{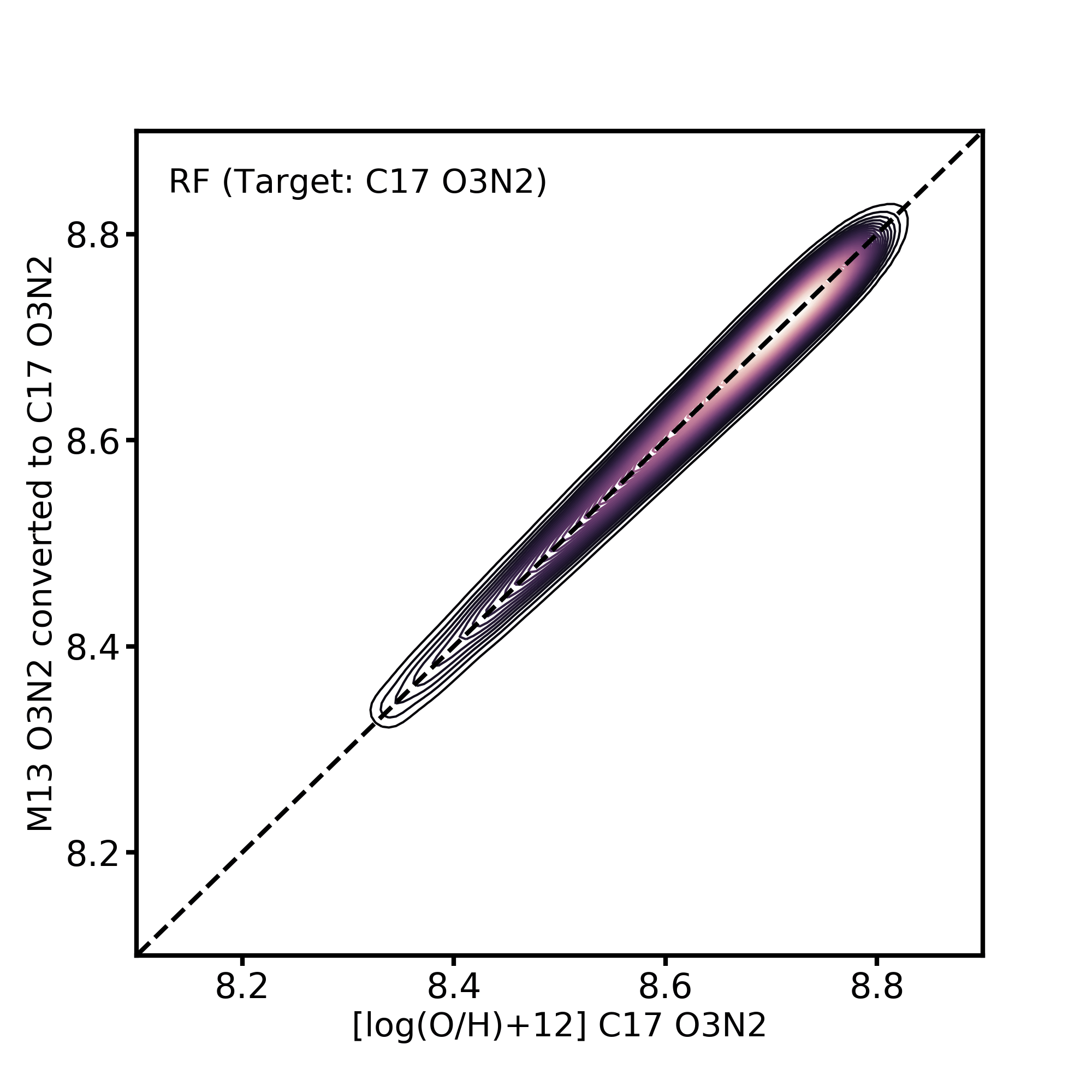}
\includegraphics[width=4.2cm]{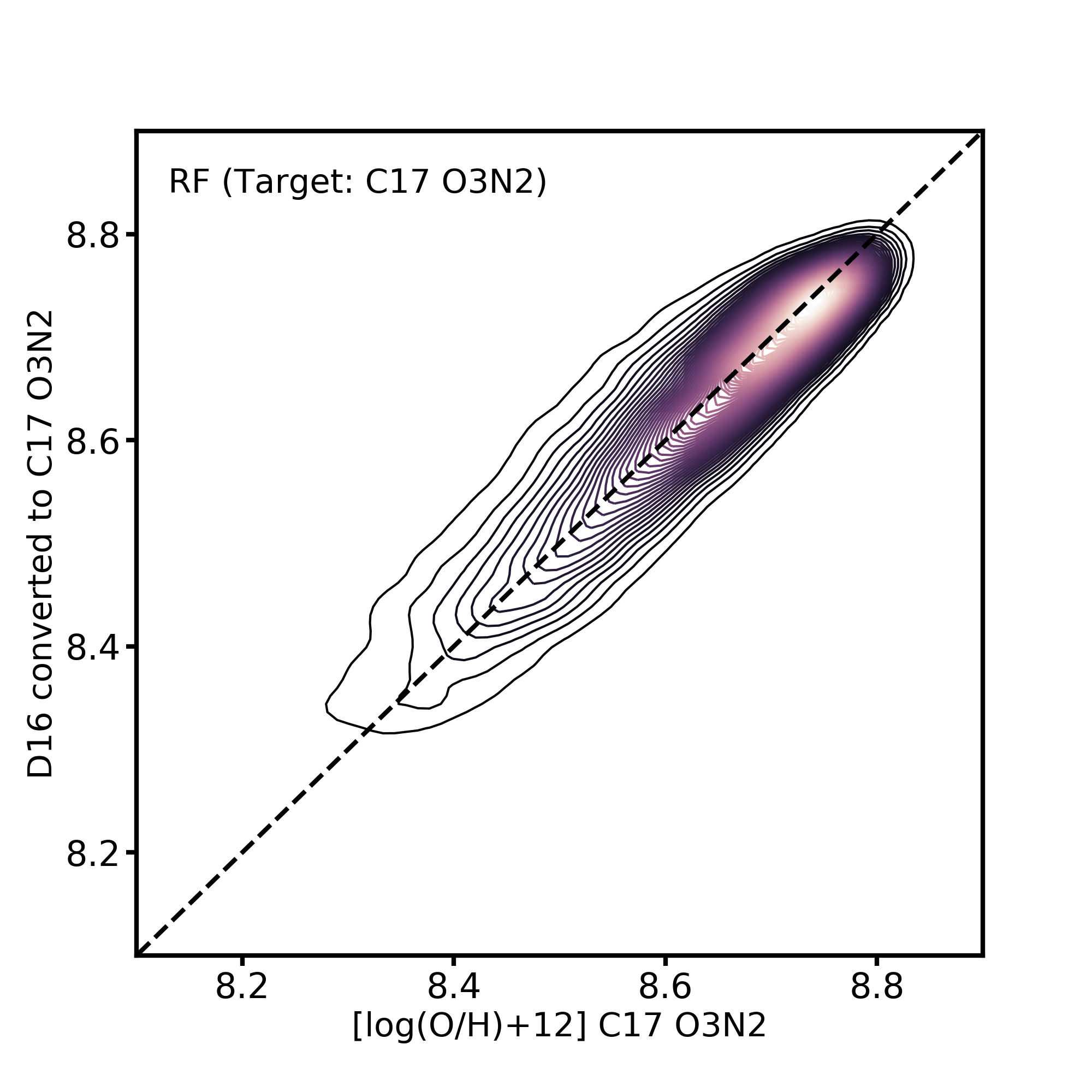}
\includegraphics[width=4.2cm]{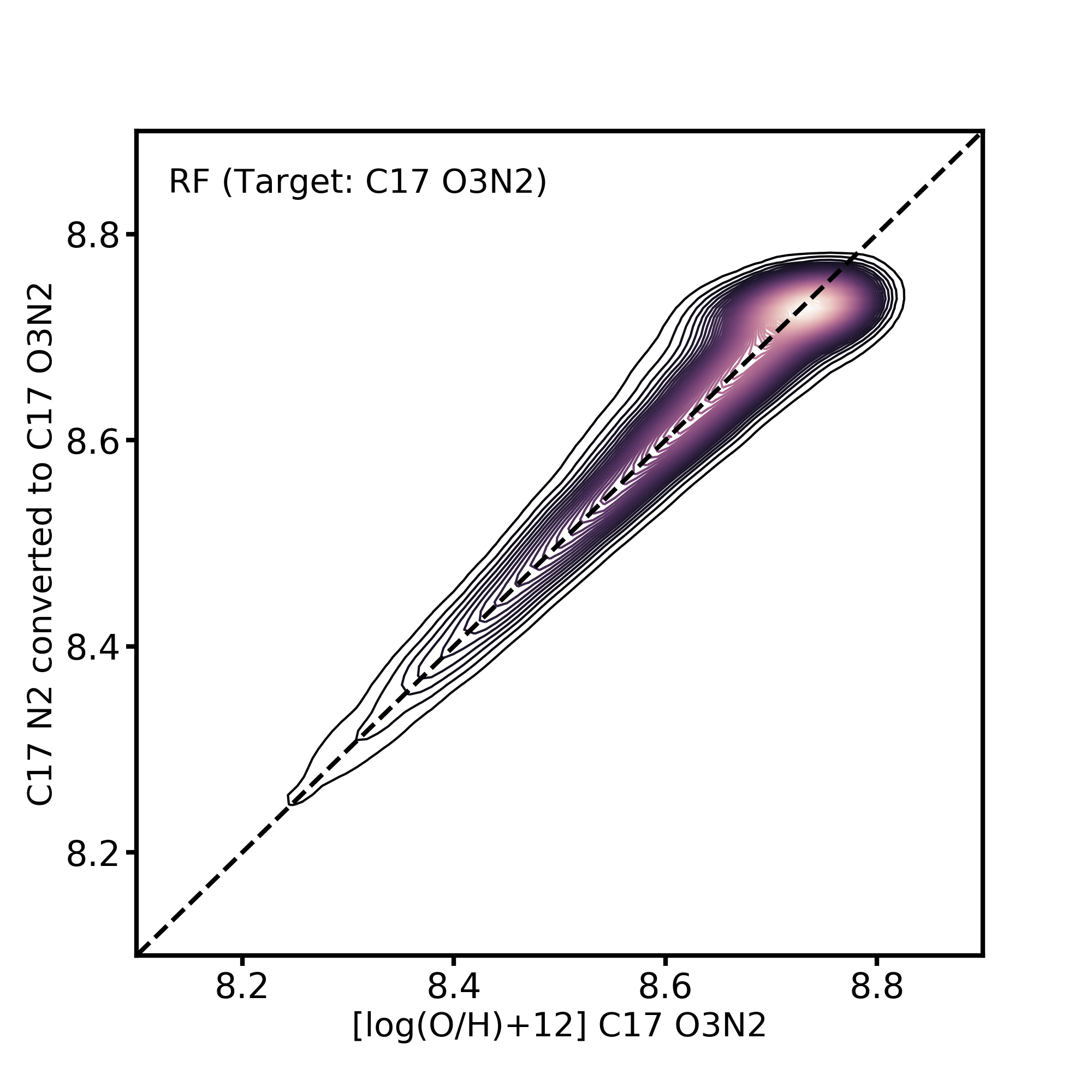}
\caption{The same as Fig. \ref{fig-C17} but using a random forest method for calculating the metallicity conversions.}
\label{fig-RF}
\end{figure*}

Fig. \ref{fig-RF} has the same format as Fig. \ref{fig-C17}, i.e. using C17 O3N2  as the target and converting from the 11 other strong line methods.  Fig. \ref{fig-RF} shows that the RF models have successfully converted from the base to the target metallicity diagnostic with no systematic offsets or skewness.

\begin{figure}
\centering
\includegraphics[width=8.5cm]{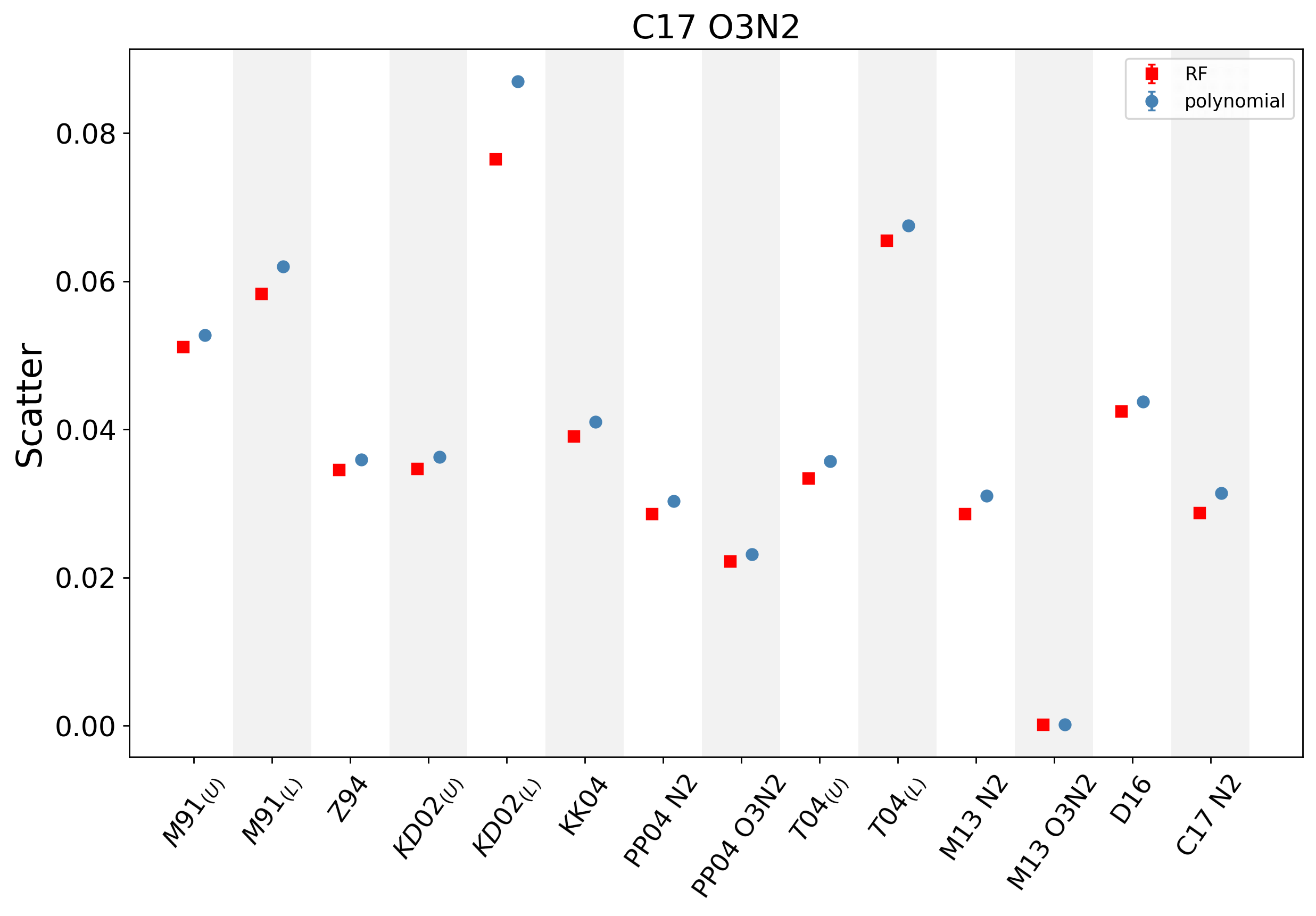}
\caption{A comparison between the scatter obtained by the new polynomial equations and the RF results. The results do not show a significant difference. The title shows the base, and on the `$x$' axis the input is shown.}
\label{fig:accuracy}
\end{figure}

Visually, Figs. \ref{fig-RF} (RF conversions with C17 O3N2 as the target) and \ref{fig-new} (DR7 polynomial fits with C17 O3N2 as the target) look very similar.   In Fig. \ref{fig:accuracy} we compare the two methods quantitatively by plotting the scatter within our new polynomial (blue circles) and  random forest (red squares) conversions.  In this example, C17 O3N2 is again the target metallicity and the scatter is shown for each of the other 11 metallicity calibrations used in the conversion.   From Fig. \ref{fig:accuracy}, it can be seen that for any given base calibration, when converting to the C17 O3N2 calibration, the scatter for the two techniques is almost identical.  We have found this to be generally true for any combination of metallicity diagnostics, indicating that in terms of overall performance, the polynomial and RF methods are equally strong.  Therefore, the random forest approach does not offer any performance advantage over the polynomial method\footnote{Because of the discrete nature of the RF `leaf', there may be more clustering around particular target metallicities in the RF model, compared to the polynomial conversion. The preservation of equal magnitudes of scatter indicates that this is not causing major biases in the conversion process.}.  However, operationally, the random forest is simpler for the user, requiring no subjective assessment of degeneracies or iterative solution for ionization parameter. We caution that this RF model has been trained and tested on the DR7 data set only, and may not transfer well to other data sets.

In order to facilitate the application of our RF model within the community, we have developed a GUI for all metallicity conversions in this paper. The interface uses 132 different RF models for the conversions. As with the polynomial GUI, the metallicities output by this RF GUI are rounded to 3 decimal places. The code can be downloaded from \href{https://drive.google.com/file/d/1Wtn9bfO5SDMGPKzQQM7lNJtsjWmKSJBg/view?usp=sharing}{(RF-link)} and uses the same format as the polynomial GUI also provided as a companion to this paper.

\begin{figure}
\centering
\includegraphics[width=8.5cm]{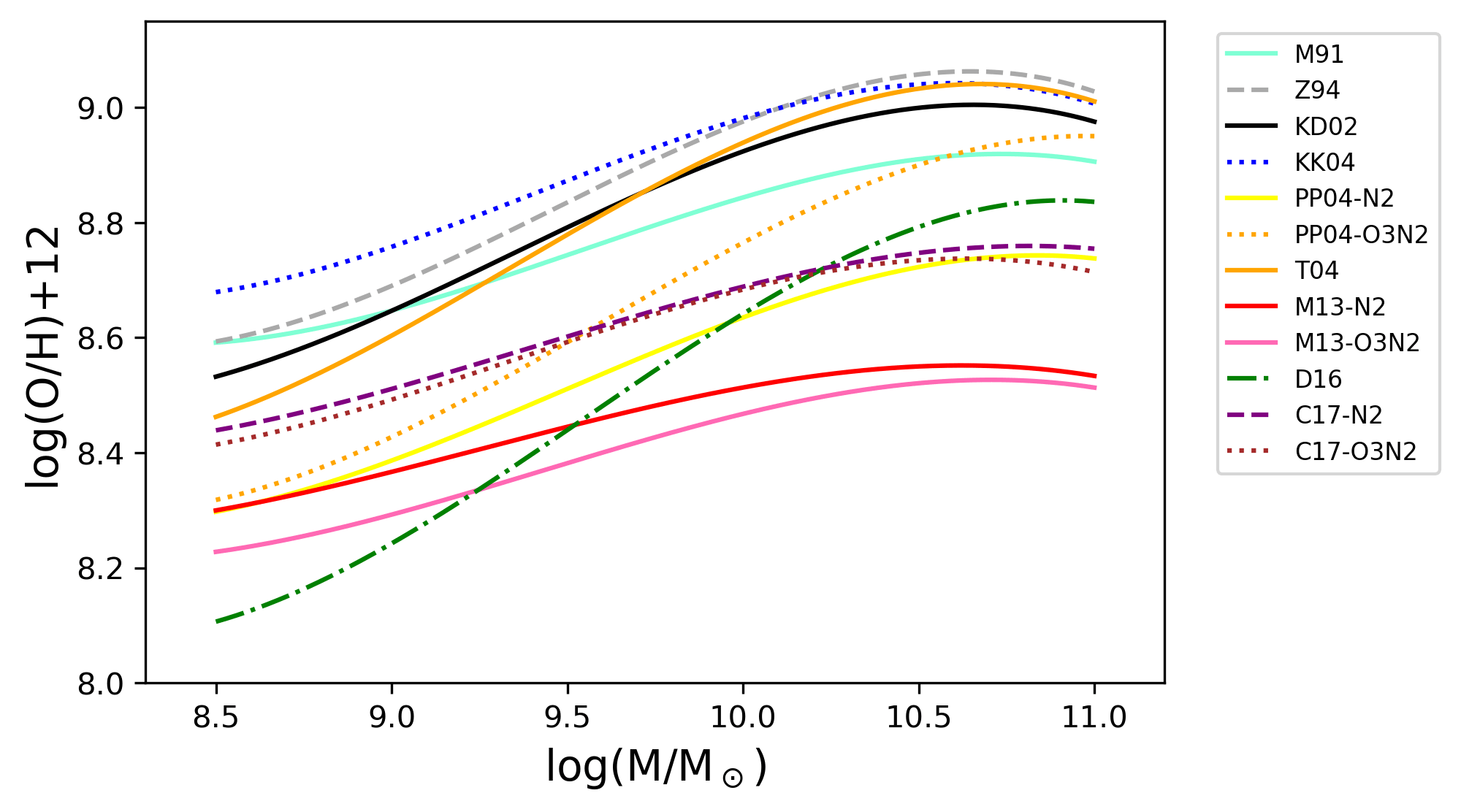}
\caption{Third order polynomial fits to the mass-metallicity relations for the original metallicity sample as recalculated for the DR7, for each of the 12 calibrations, described in Table \ref{tab:metal-ref}.}
\label{fig:metal-massNew}
\end{figure}

\begin{figure*}
\centering
\includegraphics[width=18cm,height=20.5cm,angle=0]{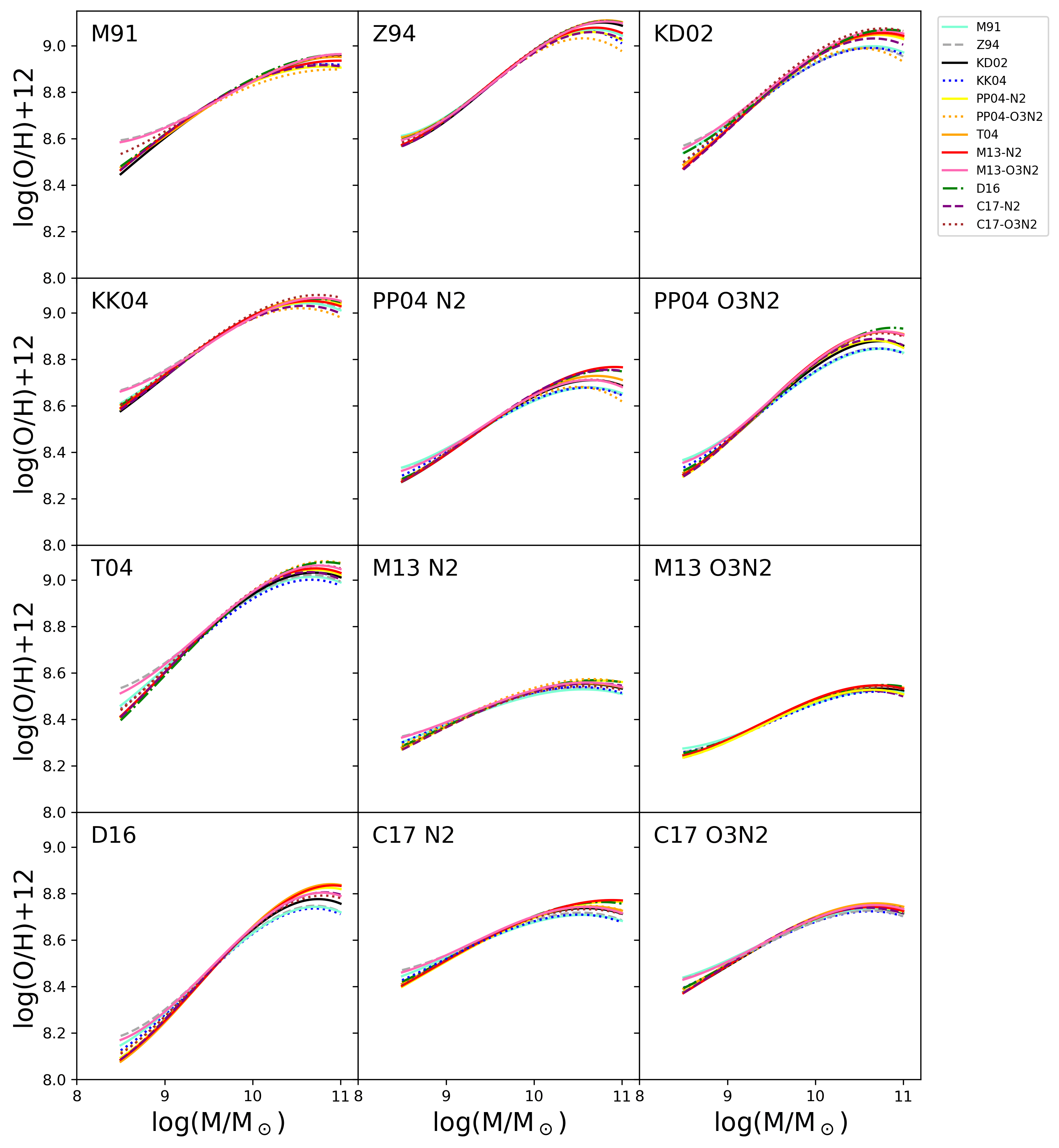}
\caption{The mass-metallicity relations derived from the DR7 data after application of the random forest conversions.  In each panel, one of the 12 metallicity calibrations is selected as the target (top left legend) and the other 11 diagnostics are converted to this target.  The curves are all in excellent agreement, showing that the conversions have successfully re-calibrated each diagnostic to a common scale.}
\label{fig:mass-metal}
\end{figure*}

Kewley \& Ellison (2008) demonstrated (their Fig. 2) that the mass-metallicity relation shows a range of 0.7 dex in metallicity (at fixed stellar mass) for the suite of calibrations used in that paper.  Kewley \& Ellison (2008) also demonstrated (their Fig. 4) that after the application of their metallicity conversions, these variations were effectively removed.  In Fig. \ref{fig:metal-massNew} and Fig. \ref{fig:mass-metal} we repeat this demonstration with the full suite of 12 calibrations used in this paper and the application of the random forest conversions.  In Fig. \ref{fig:metal-massNew}, we show a third order polynomial fit to the mass-metallicity relation for the 12 calibrations in the original metallicity sample, prior to any conversions, described in Table \ref{tab:metal-ref}.  The stellar masses come from the MPA-JHU data release of \citet{Brinchmann-04}. Broadly consistent with the results of \cite{Kewley-08}, we find a 0.6 dex range in metallicity between calibrations at fixed stellar mass.  This is slightly less than the range found by Kewley \& Ellison (2008) because we have excluded the most deviant calibration (Pilyugin \& Thuan 2005) from our suite. In Fig. \ref{fig:mass-metal}, we show the improvement to the mass-metallicity relation for the DR7 sample after the application of the RF metallicity conversions.  In each of the 12 panels, one of the metallicity calibrations is selected in turn as the target.  The other 11 diagnostics are then converted to this target using our RF model.  For each of the 12 metallicity calibrations, it can be seen that after the RF conversion has been applied, the mass-metallicity relation is invariant to the choice of metallicity diagnostic.

\section{summary}
\label{sec:summary}

We have presented a three step re-assessment of the metallicity calibration conversions originally presented in \cite{Kewley-08}.  First, we have approximately doubled the number of star-forming galaxies in the sample, by extending the original DR4 sample used by \cite{Kewley-08} to the DR7.  Our complete star-forming galaxy sample, with cuts on covering fraction and redshift, contains $\sim$61,000 galaxies and the samples for the individual metallicity calibrations (which have further cuts on emission line S/N) typically contain $\sim$40,000 galaxies.   Seven of the strong line metallicity calibrations presented in \cite{Kewley-08} are considered in this work.  We find that the metallicity conversions derived by \cite{Kewley-08} are not optimized for use on the DR7 dataset (Fig. \ref{fig-old}).  The second stage of our analysis is therefore to repeat the third order polynomial fitting procedure used by \cite{Kewley-08} to derive new coefficients (Tables \ref{tab:M91} - \ref{tab:T04}) which better represent the DR7 data (Fig. \ref{fig-new}).  We also derive coefficients for five additional strong line metallicity diagnostics that have been presented in the literature since the original work of \cite{Kewley-08} (Fig. \ref{fig-C17} and Tables \ref{tab:M13-N2} - \ref{tab:C17-O3N2}).   Taken together, the coefficients presented in Table \ref{tab:M91} - \ref{tab:C17-O3N2} represent the most complete polynomial based conversion factors for global galaxy metallicities.

The third stage of our re-assessment of metallicity conversions focuses on the potential improvement over polynomial fits that machine learning methods may provide.  Such methods can potentially better capture complexity in the data, and have the specific advantage in the realm of metallicity determinations of not requiring separate functions for upper and lower branch $R_{23}$ calibrations.  We successfully model the metallicity conversions between each of the 12 diagnostics presented in this work (Fig \ref{fig-RF}).  Despite the additional potential of the random forest compared to a simple third order polynomial fit, the two methods perform very similarly.  The predicted metallicity conversions are very similar (e.g. comparing Fig. \ref{fig-C17} and Fig. \ref{fig-RF}) and the uncertainties of the two methods are almost identical (Fig. \ref{fig:accuracy}).  Nonetheless, the RF offers the afore-mentioned advantage of avoiding uncertainty in decisions concerning upper and lower branch $R_{23}$ branches, and therefore may be more robust for metallicities in the turnover regime at intermediate metallicities.  We have made our random forest models publicly available in a user-friendly format for use by the community.

\section*{Acknowledgements}

SLE gratefully acknowledges the receipt of an NSERC Discovery Grant.

We are grateful to the MPA/JHU groups for making their SDSS catalogs public.  The work presented here would not have been possible without this resource.

Funding for the SDSS and SDSS-II has been provided by the Alfred P. Sloan Foundation, the Participating Institutions, the National Science Foundation, the U.S. Department of Energy, the National Aeronautics and Space Administration, the Japanese Monbukagakusho, the Max Planck Society, and the Higher Education Funding Council for England. The SDSS Web Site is http://www.sdss.org/.

The SDSS is managed by the Astrophysical Research Consortium for the Participating Institutions. The Participating Institutions are the American Museum of Natural History, Astrophysical Institute Potsdam, University of Basel, University of Cambridge, Case Western Reserve University, University of Chicago, Drexel University, Fermilab, the Institute for Advanced Study, the Japan Participation Group, Johns Hopkins University, the Joint Institute for Nuclear Astrophysics, the Kavli Institute for Particle Astrophysics and Cosmology, the Korean Scientist Group, the Chinese Academy of Sciences (LAMOST), Los Alamos National Laboratory, the Max-Planck-Institute for Astronomy (MPIA), the Max-Planck-Institute for Astrophysics (MPA), New Mexico State University, Ohio State University, University of Pittsburgh, University of Portsmouth, Princeton University, the United States Naval Observatory, and the University of Washington.

\section{Data availability}

We have developed two GUIs for all polynomial and RF metallicity conversions in this paper. Each interface uses 132 different set of coefficients  and models for the conversions. 
The data underlying this article are available in:

Polynomial: \href{https://drive.google.com/file/d/1Judh3GW3emjLOE6fk-sqken9yOfacDA7/view?usp=sharing}{(Polynomial-link)}.

Random Forest:  \href{https://drive.google.com/file/d/1Wtn9bfO5SDMGPKzQQM7lNJtsjWmKSJBg/view?usp=sharing}{(RF-link)}

\bibliographystyle{mnras.bst}
\bibliography{main.bib}

\appendix

\section{Polynomial coefficients from DR7 data}

As shown in Fig. \ref{fig-old}, the \cite{Kewley-08} polynomial coefficients derived from DR4 data are not optimized for the DR7 dataset used herein.  In Tables \ref{tab:M91} -- \ref{tab:T04} we provide updated coefficients for six of the metallicity diagnostics used in \cite{Kewley-08} for the DR7 dataset, using the same polynomial form, i.e.,

\begin{equation}
y= a + bx + cx^2 + dx^3
\end{equation}
y is the target calibration (i.e. what you want to convert to), and x is the metallicity that we want to convert from. The coefficients \textit{a, b, c}, and \textit{d} can be found for different targets. For example, suppose that we have an estimated metallicity (x) that has been obtained by method M91. If we want to convert it to (target) Z94, then we should use the coefficients in the first row of Table \ref{tab:Z94}.

Furthermore, in Tables  \ref{tab:M13-N2} -- \ref{tab:C17-O3N2} we present coefficients for five more recent metallicity diagnostics included in this paper that were not available to \cite{Kewley-08}.

\begin{table*}
\caption{Metallicity calibration conversion constants for different inputs (x). Target (y) = \textbf{M91}}
\begin{tabular}{lccccccc}
\hline
x & x-range & \textit{N}-Galaxies & \textit{a} & \textit{b} & \textit{c} & \textit{d} \\
\hline
$\rm{Z94}$ & $\rm8.4-9.2$ & 31443 & 86.793 & -23.7914 & 2.3092 & -0.0697\\
$\rm{KD02_{(U)}}$ & $\rm8.4-9.2$ & 31714 & 387.3987 & -123.9031 & 13.4001 & -0.4784\\
$\rm{KD02_{(L)}}$ & $\rm8.1-8.4$ & 558 & 600.32489 & -144.5812 & 8.8228 & 0\\
$\rm{KK04}$ & $\rm8.25-9.15$ & 32102 & -2193.9176 & 751.0972 & -85.452 & 3.2434\\
$\rm{PP04\ \ N2_{(U)}}$ & $\rm8.2-8.8$ & 31133 & -225.7426 & 79.8716 & -9.1515 & 0.3529\\
$\rm{PP04\ \ N2_{(L)}}$ & $\rm8.05-8.3$ & 1813 & 4.7727 & -2.4793 & 0.3550 & 0\\
$\rm{PP04\ \ O3N2_{(U)}}$ & $\rm8.2-8.9$ & 29042 & -1033.7866 & 363.4906 & -42.3049 & 1.6435\\
$\rm{PP04\ \ O3N2_{(L)}}$ & $\rm8.05-8.4$ & 3589 & -370.6417 & 90.1443 & -5.3562 & 0\\
$\rm{T04_{(U)}}$ & $\rm8.2-9.2$ & 30871 & -354.349 & 125.8661 & -14.6117 & 0.5678\\
$\rm{T04_{(L)}}$ & $\rm8.05-8.4$ & 1160 & 1582.9942 & -588.1495 & 72.992 & -3.0104\\
$\rm{M13\ \ N2_{(U)}}$ & $\rm8.2-8.7$ & 31890 & 98.523 & -22.6664 & 1.425 & 0\\
$\rm{M13\ \ N2_{(L)}}$ & $\rm8.0-8.3$ & 1809 & -15.9858 & 2.9668 & 0 & 0\\
$\rm{M13\ \ O3N2}$ & $\rm8.1-8.7$ & 31613 & 91.0036 & -20.7636 & 1.3061 & 0\\
$\rm{D16_{(U)}}$ & $\rm8.0-9.1$ & 30758 & 53.5881 & -16.7644 & 2.0224 & -0.0788\\
$\rm{D16_{(L)}}$ & $\rm7.8-8.0$ & 536 & -8.4116 & 2.0972 & 0 & 0\\
$\rm{C17\ \ N2_{(U)}}$ & $\rm8.3-8.9$ & 31470 & 95.6496 & -33.318 & 4.092 & -0.1619\\
$\rm{C17\ \ N2_{(L)}}$ & $\rm8.1-8.3$ & 257 & -3.2641 & 1.3817 & 0 & 0\\
$\rm{C17\ \ O3N2}$ & $\rm8.25-8.9$ & 32070 & 57.9884 & -12.5818 & 0.7971 & 0\\
\hline
\end{tabular}
\label{tab:M91}
\end{table*}

\begin{table*}
\caption{Metallicity calibration conversion constants for different inputs (x). Target (y) = \textbf{Z94}}
\begin{tabular}{lccccccc}
\hline
x & x-range & \textit{N}-Galaxies & \textit{a} & \textit{b} & \textit{c} & \textit{d} \\
\hline
$\rm{M91}$ & $\rm8.4-9.1$ & 31909 & 1825.2332 & -627.9727 & 72.194 & -2.7599\\
$\rm{KD02}$ & $\rm8.4-9.2$ & 31680 & 1012.4977 & -340.8346 & 38.446 & -1.4403\\
$\rm{KK04}$ & $\rm8.55-9.2$ & 31835 & -590.5284 & 203.7411 & -23.225 & 0.8876\\
$\rm{PP04\ \ N2}$ & $\rm8.05-8.8$ & 30912 & 2619.495 & -924.9867 & 109.0873 & -4.2822\\
$\rm{PP04\ \ O3N2}$ & $\rm8.05-8.9$ & 28917 & 406.0196 & -140.3262 & 16.4181 & -0.6362\\
$\rm{T04}$ & $\rm8.4-9.2$ & 29952 & 830.7654 & -280.1803 & 31.7195 & -1.1925\\
$\rm{M13\ \ N2}$ & $\rm8.1-8.7$ & 31701 & 145.2596 & -34.2738 & 2.1454 & 0\\
$\rm{M13\ \ O3N2}$ & $\rm8.15-8.7$ & 31433 & 1437.9466 & -507.8459 & 59.9503 & -2.3507\\
$\rm{D16}$ & $\rm7.9-9.1$ & 30745 & 468.8946 & -164.737 & 19.5684 & -0.7711\\
$\rm{C17\ \ N2}$ & $\rm8.2-8.9$ & 31172 & 10.5919 & -1.9581 & 0.2038 & 0\\
$\rm{C17\ \ O3N2}$ & $\rm8.1-8.9$ & 31701 & 391.6684 & -124.3462 & 13.2068 & -0.4563\\
\hline
\end{tabular}
\label{tab:Z94}
\end{table*}

\begin{table*}
\caption{Metallicity calibration conversion constants for different inputs (x). Target (y) = \textbf{KD02}}
\begin{tabular}{lccccccc}
\hline
x & x-range & \textit{N}-Galaxies & \textit{a} & \textit{b} & \textit{c} & \textit{d} \\
\hline
$\rm{M91_{(U)}}$ & $\rm8.5-9.1$ & 31560 & 234.9239 & -87.0053 & 10.8854 & -0.4451\\
$\rm{M91_{(L)}}$ & $\rm8.05-8.3$ & 382 & 69.0297 & -15.5757 & 0.9969 & 0\\
$\rm{Z94}$ & $\rm8.4-9.2$ & 31445 & -82.8695 & 27.6615 & -2.8345 & 0.0993\\
$\rm{KK04}$ & $\rm8.2-9.2$ & 32168 & 99.3232 & -31.9684 & 3.6341 & -0.1331\\
$\rm{PP04\ \ N2_{(U)}}$ & $\rm8.2-8.9$ & 37304 & 491.9975 & -178.472 & 21.7866 & -0.8797\\
$\rm{PP04\ \ N2_{(L)}}$ & $\rm8.05-8.3$ & 1839 & 335.5263 & -81.3035 & 5.0493 & 0\\
$\rm{PP04\ \ O3N2_{(U)}}$ & $\rm8.2-8.9$ & 33017 & -353.9603 & 121.8415 & -13.7148 & 0.5177\\
$\rm{PP04\ \ O3N2_{(L)}}$ & $\rm8.05-8.3$ & 1420 & 129.8919 & -30.8517 & 1.9554 & 0\\
$\rm{T04_{(U)}}$ & $\rm8.2-9.2$ & 36029 & 289.3069 & -97.0787 & 11.0951 & -0.4188\\
$\rm{T04_{(L)}}$ & $\rm8.05-8.4$ & 1186 & -139.3008 & 50.8218 & -5.9044 & 0.232\\
$\rm{M13\ \ N2_{(U)}}$ & $\rm8.2-8.7$ & 37497 & 2113.0985 & -744.0017 & 87.4392 & -3.4156\\
$\rm{M13\ \ N2_{(L)}}$ & $\rm8.0-8.2$ & 377 & 450.6873 & -109.3972 & 6.7626 & 0\\
$\rm{M13\ \ O3N2}$ & $\rm8.15-8.7$ & 37214 & -1831.2202 & 651.7518 & -77.1251 & 3.04923\\
$\rm{D16}$ & $\rm8.0-9.1$ & 35930 & -105.37 & 36.6551 & -3.962 & 0.1447\\
$\rm{C17\ \ N2_{(U)}}$ & $\rm8.3-8.9$ & 36758 & -39.973 & 9.7639 & -0.4758 & 0\\
$\rm{C17\ \ N2_{(L)}}$ & $\rm8.1-8.3$ & 258 & 29.1097 & -5.3562 & 0.3433 & 0\\
$\rm{C17\ \ O3N2_{(U)}}$ & $\rm8.25-8.9$ & 31992 & -770.9174 & 271.901 & -31.7625 & 1.2428\\
$\rm{C17\ \ O3N2_{(L)}}$ & $\rm8.0-8.25$ & 174 & -61.3755 & 16.8597 & -1.0199 & 0\\
\hline
\end{tabular}
\label{tab:KD02}
\end{table*}

\begin{table*}
\caption{Metallicity calibration conversion constants for different inputs (x). Target (y) = \textbf{KK04}}
\begin{tabular}{lccccccc}
\hline
x & x-range & \textit{N}-Galaxies & \textit{a} & \textit{b} & \textit{c} & \textit{d} \\
\hline
$\rm{M91_{(U)}}$ & $\rm8.4-9.1$ & 31919 & 1396.2622 & -483.698 & 56.06155 & -2.1602\\
$\rm{M91_{(L)}}$ & $\rm8.0-8.25$ & 425 & 721.01897 & -260.4697 & 31.681 & -1.28229\\
$\rm{Z94}$ & $\rm8.4-9.3$ & 31894 & 204.83638 & -69.615722 & 8.13631 & -0.31321\\
$\rm{KD02_{(U)}}$ & $\rm8.4-9.2$ & 31689 & 1146.076 & -388.5092 & 44.1343 & -1.6671\\
$\rm{KD02_{(L)}}$ & $\rm8.05-8.3$ & 356 & 428.7191 & -153.4709 & 18.5653 & -0.74416\\
$\rm{PP04\ \ N2_{(U)}}$ & $\rm8.2-8.9$ & 31715 & 1137.2633 & -404.9428 & 48.3076 & -1.9159\\
$\rm{PP04\ \ N2_{(L)}}$ & $\rm8.05-8.3$ & 1718 & 221.1257 & -53.9172 & 3.4127 & 0\\
$\rm{PP04\ \ O3N2_{(U)}}$ & $\rm8.2-8.9$ & 28959 & -172.8182 & 60.3876 & -6.7529 & 0.2543\\
$\rm{PP04\ \ O3N2_{(L)}}$ & $\rm8.05-8.3$ & 1343 & 37.2557 & -8.5021 & 0.6094 & 0\\
$\rm{T04_{(U)}}$ & $\rm8.3-9.2$ & 30407 & 889.9372 & -301.6969 & 34.3416 & -1.2994\\
$\rm{T04_{(L)}}$ & $\rm8.05-8.4$ & 1071 & 3967.1747 & -1449.9755 & 176.8893 & -7.1874\\
$\rm{M13\ \ N2_{(U)}}$ & $\rm8.2-8.7$ & 31813 & 3853.3363 & -1367.5488 & 161.9577 & -6.3857\\
$\rm{M13\ \ N2_{(L)}}$ & $\rm8.0-8.2$ & 345 & 610.3371 & -149.0526 & 9.2261 & 0\\
$\rm{M13\ \ O3N2}$ & $\rm8.15-8.7$ & 31550 & 542.1259 & -193.5488 & 23.2532 & -0.9248\\
$\rm{D16_{(U)}}$ & $\rm8.0-9.1$ & 30695 & 259.97912 & -91.48058 & 11.02932 & -0.44022\\
$\rm{D16_{(L)}}$ & $\rm7.8-8.0$ & 494 & -1.86835 & 1.29895 & 0 & 0\\
$\rm{C17\ \ N2_{(U)}}$ & $\rm8.3-8.9$ & 31385 & 3705.0113 & -1298.9898 & 152.0019 & -5.9222\\
$\rm{C17\ \ N2_{(L)}}$ & $\rm8.1-8.3$ & 238 & 4.148 & 0.5126 & 0 & 0\\
$\rm{C17\ \ O3N2_{(U)}}$ & $\rm8.3-8.9$ & 31749 & -374.6472 & 136.3397 & -16.2909 & 0.6538\\
$\rm{C17\ \ O3N2_{(L)}}$ & $\rm8.1-8.3$ & 400 & 293.2935 & -70.2175 & 4.3258 & 0\\
\hline
\end{tabular}
\label{tab:KK04}
\end{table*}

\begin{table*}
\caption{Metallicity calibration conversion constants for different inputs (x). Target (y) = \textbf{PP04 N2}}
\begin{tabular}{lccccccc}
\hline
x & x-range & \textit{N}-Galaxies & \textit{a} & \textit{b} & \textit{c} & \textit{d}\\
\hline
$\rm{M91_{(U)}}$ & $\rm8.5-9.1$ & 31431 & 3613.3844 & -1238.8022 & 141.7739 & -5.4034\\
$\rm{M91_{(L)}}$ & $\rm8.05-8.4$ & 407 & 163.9517 & -38.6531 & 2.398 & 0\\
$\rm{Z94}$ & $\rm8.4-9.3$ & 31540 & 856.0385 & -290.6646 & 33.1364 & -1.2557\\
$\rm{KD02}$ & $\rm8.05-9.2$ & 37572 & 865.4984 & -295.5199 & 33.8635 & -1.2895\\
$\rm{KK04}$ & $\rm8.2-9.2$ & 32039 & 733.3485 & -246.8248 & 27.9109 & -1.0481\\
$\rm{PP04\ \ O3N2}$ & $\rm8.05-8.9$ & 38278 & 409.144 & -143.4804 & 17.0264 & -0.6696\\
$\rm{T04}$ & $\rm8.05-9.2$ & 40978 & 524.796 & -178.1117 & 20.3914 & -0.7747\\
$\rm{M13\ \ N2}$ & $\rm8.0-8.6$ & 44686 & 163.1399 & -38.4256 & 2.3818 & 0\\
$\rm{M13\ \ O3N2}$ & $\rm8.1-8.7$ & 43759 & 3794.7691 & -1361.791 & 163.0805 & -6.5026\\
$\rm{D16}$ & $\rm7.8-9.1$ & 42591 & 141.2938 & -49.2383 & 5.9996 & -0.2404\\
$\rm{C17\ \ N2}$ & $\rm8.0-8.9$ & 43741 & 1229.0742 & -424.8038 & 49.1162 & -1.8864\\
$\rm{C17\ \ O3N2}$ & $\rm8.0-8.9$ & 44611 & 1196.6207 & -416.474 & 48.5111 & -1.8776\\
\hline
\end{tabular}
\label{tab:PP04-N2}
\end{table*}

\begin{table*}
\caption{Metallicity calibration conversion constants for different inputs (x). Target (y) = \textbf{PP04 O3N2}}
\begin{tabular}{lccccccc}
\hline
x & x-range & \textit{N}-Galaxies & \textit{a} & \textit{b} & \textit{c} & \textit{d} \\
\hline
$\rm{M91_{(U)}}$ & $\rm8.5-9.1$ & 31555 & 2855.1815 & -976.6748 & 111.5376 & -4.2396\\
$\rm{M91_{(L)}}$ & $\rm 8.05-8.4$ & 387 & -1545.1399 & 587.9352 & -74.1884 & 3.1208\\
$\rm{Z94}$ & $\rm8.4-9.3$ & 31540 & 342.0709 & -114.2818 & 12.942 & -0.4843\\
$\rm{KD02}$ & $\rm8.1-9.2$ & 37672 & 1005.4986 & -342.3498 & 39.0563 & -1.4802\\
$\rm{KK04}$ & $\rm8.2-9.2$ & 32100 & 263.38 & -82.1518 & 8.6714 & -0.2985\\
$\rm{PP04\ \ N2}$ & $\rm8.05-8.8$ & 42677 & 1733.7386 & -614.419 & 72.766 & -2.8657\\
$\rm{T04}$ & $\rm8.05-9.2$ & 40976 & 425.2876 & -141.5692 & 15.9152 & -0.5918\\
$\rm{M13\ \ N2}$ & $\rm8.1-8.7$ & 44584 & 155.631 & -37.0791 & 2.329 & 0\\
$\rm{M13\ \ O3N2}$ & $\rm8.15-8.7$ & 43840 & 197.388 & -65.8684 & 7.4331 & -0.2698\\
$\rm{D16}$ & $\rm7.8-9.1$ & 42611 & 237.6059 & -83.634 & 10.0731 & -0.4003\\
$\rm{C17\ \ N2}$ & $\rm8.1-8.9$ & 43289 & 5361.8221 & -1875.2013 & 218.7172 & -8.4942\\
$\rm{C17\ \ O3N2}$ & $\rm8.1-8.9$ & 44586 & -185.4252 & 76.7659 & -10.1843 & 0.4513\\
\hline
\end{tabular}
\label{tab:PP04-O3N2}
\end{table*}

\begin{table*}
\caption{Metallicity calibration conversion constants for different inputs (x). Target (y) = \textbf{T04}}
\begin{tabular}{lccccccc}
\hline
x & x-range & \textit{N}-Galaxies & \textit{a} & \textit{b} & \textit{c} & \textit{d} \\
\hline
$\rm{M91_{(U)}}$ & $\rm8.4-9.2$ & 30651 & 2998.9561 & -1030.2075 & 118.139 & -4.5092\\
$\rm{M91_{(L)}}$ & $\rm8.0-8.4$ & 500 & 2.0925 & 0.7576 & 0 & 0\\
$\rm{Z94}$ & $\rm8.4-9.3$ & 30155 & 549.9891 & -185.833 & 21.1524 & -0.7982\\
$\rm{KD02_{(U)}}$ & $\rm8.4-9.3$ & 35852 & 521.5475 & -179.6248 & 20.834 & -0.8003\\
$\rm{KD02_{(L)}}$ & $\rm8.1-8.4$ & 559 & -2.4905 & 1.29317 & 0 & 0\\
$\rm{KK04_{(U)}}$ & $\rm8.45-9.2$ & 30654 & 1306.7925 & -437.9528 & 49.1096 & -1.8301\\
$\rm{KK04_{(L)}}$ & $\rm8.2-8.4$ & 457 & -6.0523 & 1.7088 & 0 & 0\\
$\rm{PP04\ \ N2_{(U)}}$ & $\rm8.4-8.9$ & 35781 & -104.9897 & 25.3186 & -1.4041 & 0\\
$\rm{PP04\ \ N2_{(L)}}$ & $\rm8.0-8.4$ & 5267 & -6.825 & 1.8446 & 0 & 0\\
$\rm{PP04\ \ O3N2_{(U)}}$ & $\rm8.4-9.2$ & 37478 & 522.7574 & -181.3092 & 21.2074 & -0.8225\\
$\rm{PP04\ \ O3N2_{(L)}}$ & $\rm8.0-8.4$ & 3600 & 947.9234 & -349.4912 & 43.1432 & -1.7678\\
$\rm{M13\ \ N2}$ & $\rm8.0-8.7$ & 40624 & -804.0024 & -279.9536 & 32.5884 & -1.2538\\
$\rm{M13\ \ O3N2}$ & $\rm8.1-8.7$ & 39970 & -1821.5398 & 651.9099 & -77.5969 & 3.0867\\
$\rm{D16_{(U)}}$ & $\rm8.1-9.1$ & 39432 & 244.9229 & -85.9962 & 10.3362 & -0.4102\\
$\rm{D16_{(L)}}$ & $\rm7.8-8.1$ & 1608 & 581.3903 & -188.2034 & 19.9897 & -0.6774\\
$\rm{C17\ \ N2}$ & $\rm8.2-8.9$ & 39543 & -43.2538 & 10.3962 & -0.5052 & 0\\
$\rm{C17\ \ O3N2}$ & $\rm8.2-8.9$ & 40634 & -1296.48795 & 461.04 & -54.4573 & 2.1507\\
\hline
\end{tabular}
\label{tab:T04}
\end{table*}

\begin{table*}
\caption{Metallicity calibration conversion constants for different inputs (x). Target (y) = \textbf{M13 N2}}
\begin{tabular}{lccccccc}
\hline
x & x-range & \textit{N}-Galaxies & \textit{a} & \textit{b} & \textit{c} & \textit{d} \\
\hline
$\rm{M91_{(U)}}$ & $\rm8.5-9.1$ & 31560 & 1559.59 & -537.0831 & 61.9093 & -2.3757\\
$\rm{M91_{(L)}}$ & $\rm7.9-8.4$ & 562 & 48.3787 & -10.4483 & 0.6772 & 0\\
$\rm{Z94}$ & $\rm8.4-9.3$ & 31669 & 226.7235 & -77.9489 & 9.2069 & -0.3599\\
$\rm{KD02_{(U)}}$ & $\rm8.4-9.3$ & 37326 & 380.7872 & -131.0384 & 15.2996 & -0.5928\\
$\rm{KD02_{(L)}}$ & $\rm8.2-8.4$ & 528 & -0.7239 & 1.6589 & -0.0704 & 0\\
$\rm{KK04_{(U)}}$ & $\rm8.55-9.2$ & 31612 & 805.682 & -273.9803 & 31.3139 & -1.1903\\
$\rm{KK04_{(L)}}$ & $\rm8.2-8.4$ & 447 & 4.352 & 0.4586 & 0 & 0\\
$\rm{PP04\ \ N2}$ & $\rm8.0-8.9$ & 44709 & -460.0623 & 158.8308 & -17.9864 & 0.6805\\
$\rm{PP04\ \ O3N2}$ & $\rm8.0-9.2$ & 44585 & -86.985 & 28.6534 & -2.8417 & 0.093\\
$\rm{T04_{(U)}}$ & $\rm8.55-9.2$ & 38918 & 0.3458 & -1.8454 & 0.6829 & -0.0418\\
$\rm{T04_{(L)}}$ & $\rm8.0-8.5$ & 2020 & 574.6886 & -212.1147 & 26.3888 & -1.0909\\
$\rm{M13\ \ O3N2}$ & $\rm8.1-8.7$ & 44085 & 120.9708 & -52.0238 & 7.5124 & -0.3468\\
$\rm{D16}$ & $\rm7.8-9.1$ & 42863 & -64.0936 & 22.8361 & -2.3987 & 0.0842\\
$\rm{C17\ \ N2}$ & $\rm8.0-8.9$ & 43734 & 932.1046 & -329.5358 & 39.0721 & -1.5398\\
$\rm{C17\ \ O3N2}$ & $\rm8.0-8.9$ & 44929 & 697.0506 & -247.6338 & 29.572 & -1.1729\\
\hline
\end{tabular}
\label{tab:M13-N2}
\end{table*}

\begin{table*}
\caption{Metallicity calibration conversion constants for different inputs (x). Target (y) = \textbf{M13 O3N2}}
\begin{tabular}{lccccccc}
\hline
x & x-range & \textit{N}-Galaxies & \textit{a} & \textit{b} & \textit{c} & \textit{d} \\
\hline
$\rm{M91_{(U)}}$ & $\rm8.5-9.1$ & 31282 & 1888.2496 & -644.077 & 73.4676 & -2.7899\\
$\rm{M91_{(L)}}$ & $\rm8.0-8.3$ & 167 & -1305.8981 & 491.485 & -61.2792 & 2.547\\
$\rm{Z94}$ & $\rm8.4-9.3$ & 31417 & 341.7221 & -113.7254 & 12.8692 & -0.483\\
$\rm{KD02}$ & $\rm8.4-9.3$ & 37043 & 967.8764 & -328.278 & 37.3634 & -1.4146\\
$\rm{KK04_{(U)}}$ & $\rm8.55-9.2$ & 31362 & 536.8013 & -176.7522 & 19.6262 & -0.7232\\
$\rm{KK04_{(L)}}$ & $\rm8.2-8.4$ & 116 & 98.1808 & -21.7183 & 1.3102 & 0\\
$\rm{PP04\ \ N2}$ & $\rm8.2-8.9$ & 43757 & 1204.6286 & -425.8859 & 50.4443 & -1.9879\\
$\rm{PP04\ \ O3N2}$ & $\rm8.2-9.1$ & 43707 & 234.8163 & -81.3845 & 9.6777 & -0.3809\\
$\rm{T04_{(U)}}$ & $\rm8.55-9.2$ & 38815 & 182.9739 & -61.21 & 7.0837 & -0.2707\\
$\rm{T04_{(L)}}$ & $\rm8.1-8.5$ & 1477 & 2411.7063 & -867.0572 & 104.2289 & -4.175\\
$\rm{M13\ \ N2}$ & $\rm8.2-8.65$ & 44082 & 6083.3786 & -2158.8367 & 255.5638 & -10.0782\\
$\rm{D16}$ & $\rm7.8-9.1$ & 42012 & 214.9474 & -74.7581 & 8.962 & -0.3559\\
$\rm{C17\ \ N2}$ & $\rm8.3-8.9$ & 42782 & 5320.2935 & -1857.9627 & 216.4846 & -8.4027\\
$\rm{C17\ \ O3N2}$ & $\rm8.3-8.9$ & 44083 & 0.3628 & 0.9335 & 0 & 0\\
\hline
\end{tabular}
\label{tab:M13-O3N2}
\end{table*}

\begin{table*}
\caption{Metallicity calibration conversion constants for different inputs (x). Target (y) = \textbf{D16}}
\begin{tabular}{lccccccc}
\hline
x & x-range & \textit{N}-Galaxies & \textit{a} & \textit{b} & \textit{c} & \textit{d} \\
\hline
$\rm{M91_{(U)}}$ & $\rm8.5-9.1$ & 30613 & 2541.5169 & -873.1917 & 100.1378 & -3.8204\\
$\rm{M91_{(L)}}$ & $\rm8.0-8.3$ & 471 & 5.1048 & 0.3513 & 0 & 0\\
$\rm{Z94}$ & $\rm8.4-9.3$ & 30224 & 3706.8072 & -237.5519 & 26.8035 & -1.0031\\
$\rm{KD02_{(U)}}$ & $\rm8.4-9.3$ & 35949 & 1292.1462 & -439.7038 & 50.0465 & -1.8928\\
$\rm{KD02_{(L)}}$ & $\rm8.2-8.4$ & 524 & 3.3878 & 0.5506 & 0 & 0\\
$\rm{KK04_{(U)}}$ & $\rm8.55-9.2$ & 30663 & 875.0474 & -289.4371 & 32.0548 & -1.1768\\
$\rm{KK04_{(L)}}$ & $\rm8.2-8.4$ & 443 & 2.8759 & 0.6069 & 0 & 0\\
$\rm{PP04\ \ N2}$ & $\rm8.1-8.9$ & 41901 & 957.6926 & -337.1474 & 39.7259 & -1.5529\\
$\rm{PP04\ \ O3N2}$ & $\rm8.0-9.2$ & 41944 & 284.8894 & -99.0825 & 11.6981 & -0.4551\\
$\rm{T04_{(U)}}$ & $\rm8.5-9.3$ & 39034 & -252.7935 & 87.9578 & -10.0078 & 0.3848\\
$\rm{T04_{(L)}}$ & $\rm8.0-8.5$ & 2015 & 1028.8981 & -371.8225 & 45.0596 & -1.8169\\
$\rm{M13\ \ N2}$ & $\rm8.0-8.7$ & 42210 & -1633.1334 & 615.017 & -76.9143 & 3.2096\\
$\rm{M13\ \ O3N2}$ & $\rm8.15-8.65$ & 41348 & -718.1404 & 264.3618 & -32.295 & 1.3239\\
$\rm{C17\ \ N2}$ & $\rm8.0-8.9$ & 41006 & 2375.3459 & -823.6716 & 95.2887 & -3.6646\\
$\rm{C17\ \ O3N2}$ & $\rm8.0-8.9$ & 42175 & -482.1501 & 183.9562 & -23.1037 & 0.9705\\
\hline
\end{tabular}
\label{tab:D16}
\end{table*}

\begin{table*}
\caption{Metallicity calibration conversion constants for different inputs (x). Target (y) = \textbf{C17 N2}}
\begin{tabular}{lccccccc}
\hline
x & x-range & \textit{N}-Galaxies & \textit{a} & \textit{b} & \textit{c} & \textit{d} \\
\hline
$\rm{M91_{(U)}}$ & $\rm8.4-9.1$ & 31170 & 2901.7049 & -994.8047 & 113.9298 & -4.3457\\
$\rm{M91_{(L)}}$ & $\rm8.0-8.4$ & 502 & 2.5221 & 0.7141 & 0 & 0\\
$\rm{Z94}$ & $\rm8.4-9.3$ & 30636 & 383.5379 & -130.358 & 15.0295 & -0.5747\\
$\rm{KD02_{(U)}}$ & $\rm8.4-9.2$ & 36412 & 642.9785 & -220.0287 & 25.3594 & -0.9713\\
$\rm{KD02_{(L)}}$ & $\rm8.2-8.4$ & 531 & -4249.5242 & 1535.7596 & -184.7107 & 7.4079\\
$\rm{KK04_{(U)}}$ & $\rm8.55-9.2$ & 31083 & 986.5064 & -334.2708 & 38.0059 & -1.4373\\
$\rm{KK04_{(L)}}$ & $\rm8.2-8.4$ & 452 & 3.0422 & 0.6294 & 0 & 0\\
$\rm{PP04\ \ N2}$ & $\rm8.0-8.9$ & 43044 & -1175.2954 & 413.3419 & -48.1757 & 1.8745\\
$\rm{PP04\ \ O3N2}$ & $\rm8.0-9.2$ & 42596 & -7.9268 & 1.0615 & 0.3627 & -0.0305\\
$\rm{T04_{(U)}}$ & $\rm8.55-9.2$ & 37884 & 371.1892 & -127.1447 & 14.7861 & -0.5704\\
$\rm{T04_{(L)}}$ & $\rm8.0-8.5$ & 2024 & 230.5072 & -90.1699 & 11.9911 & -0.5242\\
$\rm{M13\ \ N2}$ & $\rm8.0-8.6$ & 43036 & -2591.8417 & 935.0384 & -112.197 & 4.4924\\
$\rm{M13\ \ O3N2}$ & $\rm8.15-8.65$ & 42099 & 1256.816 & -457.1278 & 55.6542 & -2.2528\\
$\rm{D16}$ & $\rm7.8-9.1$ & 40982 & -83.8363 & 30.1166 & -3.295 & 0.1213\\
$\rm{C17\ \ O3N2}$ & $\rm8.0-8.9$ & 42941 & 445.5391 & -158.2197 & 18.9688 & -0.7533\\
\hline
\end{tabular}
\label{tab:C17-N2}
\end{table*}

\begin{table*}
\caption{Metallicity calibration conversion constants for different inputs (x). Target (y) = \textbf{C17 O3N2}}
\begin{tabular}{lccccccc}
\hline
x & x-range & \textit{N}-Galaxies & \textit{a} & \textit{b} & \textit{c} & \textit{d} \\
\hline
$\rm{M91_{(U)}}$ & $\rm8.5-9.1$ & 31560 & 1922.6664 & -659.1048 & 75.5519 & -2.8829\\
$\rm{M91_{(L)}}$ & $\rm8.0-8.3$ & 477 & -31.2375 & 9.0558 & -0.5155 & 0\\
$\rm{Z94}$ & $\rm8.4-9.3$ & 31669 & 73.9083 & -25.0406 & 3.0904 & -0.1237\\
$\rm{KD02_{(U)}}$ & $\rm8.4-9.2$ & 37208 & 719.4539 & -245.8567 & 28.2585 & -1.0795\\
$\rm{KD02_{(L)}}$ & $\rm8.1-8.4$ & 559 & -100.2682 & 25.5497 & -1.5024 & 0\\
$\rm{KK04}$ & $\rm8.2-9.2$ & 32168 & 302.3792 & -100.4709 & 11.3665 & -0.4254\\
$\rm{PP04\ \ N2}$ & $\rm8.05-8.9$ & 44599 & 302.2795 & -112.1483 & 14.0962 & -0.5843\\
$\rm{PP04\ \ O3N2}$ & $\rm8.05-9.2$ & 44583 & 7.1195 & -4.0341 & 0.9299 & -0.0512\\
$\rm{T04_{(U)}}$ & $\rm8.55-9.2$ & 38918 & -86.437 & 28.1687 & -2.7891 & 0.0926\\
$\rm{T04_{(L)}}$ & $\rm8.0-8.5$ & 2025 & 431.9057 & -164.2151 & 21.06348 & -0.8947\\
$\rm{M13\ \ N2}$ & $\rm8.0-8.65$ & 44930 & 1479.4498 & -528.8548 & 63.2182 & -2.5125\\
$\rm{M13\ \ O3N2}$ & $\rm8.1-8.7$ & 44083 & -464.0953 & 161.6573 & -18.517 & 0.7109\\
$\rm{D16}$ & $\rm7.8-9.1$ & 42829 & 29.5937 & -10.6313 & 1.5776 & -0.0725\\
$\rm{C17\ \ N2}$ & $\rm8.0-8.9$ & 43636 & 1860.3708 & -657.1819 & 77.5899 & -3.0477\\
\hline
\end{tabular}
\label{tab:C17-O3N2}
\end{table*}

\end{document}